 \documentclass[journal]{IEEEtran}

\usepackage{cite}
\usepackage{amsmath}
\usepackage{array}
\usepackage{booktabs}
\usepackage{graphicx}
\usepackage{float}
\usepackage{subfig}
\usepackage{color}
\usepackage{amsfonts}
\usepackage{yhmath}
\usepackage{bm}
\usepackage{longtable}

\title{Compressive Sensing-Based Grant-Free Massive Access for 6G Massive Communication}

\author{Zhen Gao, Malong Ke, Yikun Mei, Li Qiao, Sheng Chen,~\IEEEmembership{Life Fellow,~IEEE},\\ 
Derrick Wing Kwan Ng,~\IEEEmembership{Fellow,~IEEE}, and H. Vincent Poor,~\IEEEmembership{Life Fellow,~IEEE}
\thanks{In this work, Zhen Gao is supported in part by the National Natural Science Foundation of China (NSFC) under Grant U2233216 and Grant 62071044, in part by the Shandong Province Natural Science Foundation under Grant ZR2022YQ62, and in part by the Beijing Nova Program; Derrick Wing Kwan Ng is supported by the Australian Research Council's Discovery Projects under Grants DP210102169 and DP230100603; H. Vincent Poor is supported by the U.S National Science Foundation under Grants CNS-2128448 and ECCS-2335876. (\textit{Corresponding author: Malong Ke})}
\thanks{Zhen Gao is with the MIIT Key Laboratory of Complex-Field Intelligent Sensing, Beijing Institute of Technology, Beijing 100081, China, and with the Yangtze Delta Region Academy, Beijing Institute of Technology (Jiaxing), Jiaxing 314019, China, and also with the Advanced Technology Research Institute, Beijing Institute of Technology, Jinan 250307, China (e-mail: gaozhen16@bit.edu.cn).
} %
\thanks{Malong Ke is with the Wireless Product Division, Ruijie Network Co. Ltd., Fuzhou 350108, China (e-mail: kemalong@ruijie.com.cn).}
\thanks{Yikun Mei and Li Qiao are with the School of Information and Electronics, Beijing Institute of Technology, 100081 Beijing, China (e-mails: meiyikun@bit.edu.cn, qiaoli@bit.edu.cn).}
\thanks{Sheng Chen is with the School of Electronics and Computer Science, University of Southampton, Southampton SO17 1BJ, U.K. (e-mail: sqc@soton.ac.uk).} %
\thanks{Derrick Wing Kwan Ng is with the School of Electrical Engineering and Telecommunications, University of New South Wales, 2052 Sydney, Australia (e-mail: w.k.ng@unsw.edu.au).} %
\thanks{H. Vincent Poor is with the Department of Electrical and Computer Engineering, Princeton University, NJ 08542 Princeton, USA (e-mail: poor@princeton.edu).}} %

\begin{document}

\maketitle

\begin{abstract}
The advent of the sixth-generation (6G) of wireless communications has given rise to the necessity to connect vast quantities of heterogeneous wireless devices, which requires advanced system capabilities far beyond existing network architectures. In particular, such massive communication has been recognized as a prime driver that can empower the 6G vision of future ubiquitous connectivity, supporting Internet of Human-Machine-Things for which massive access is critical. This paper surveys the most recent advances toward massive access in both academic and industry communities, focusing primarily on the promising compressive sensing-based grant-free massive access paradigm. We first specify the limitations of existing random access schemes and reveal that the practical implementation of massive communication relies on a dramatically different random access paradigm from the current ones mainly designed for human-centric communications. Then, a compressive sensing-based grant-free massive access roadmap is presented, where the evolutions from single-antenna to large-scale antenna array-based base stations, from single-station to cooperative massive multiple-input multiple-output systems, and from unsourced to sourced random access scenarios are detailed. Finally, we discuss the key challenges and open issues to shed light on the potential future research directions of grant-free massive access.
\end{abstract}

\begin{IEEEkeywords}
Internet-of-Things (IoT), 6G, massive communication, grant-free massive access, compressive sensing, Internet of Human-Machine-Things.
\end{IEEEkeywords}

\vspace*{5mm}
\section{Introduction}\label{S1}
In the future sixth-generation (6G) mobile network vision, the concept of Internet-of-Things (IoT) is gradually evolving into the Internet of Human-Machine-Things (IoHMT) paradigm, where the interactions across humans, machines, and things are intricately interconnected to create an intelligent ecosystem \cite{XHYou2021_6G, BinGuo2022HMT, Vermesan2019H2M, Zyang2021HMT}. In the upcoming IoHMT era, the ubiquitous connectivity of heterogeneous devices is expected to enable a plethora of promising applications, such as smart cities and smart factories, promoting the digitalization of society, and improving the overall efficiency of various vertical sectors \cite{Mahmood2020mMTC}. It is predicted that there will be up to 75 billion devices connected in IoHMT ecosystems by 2025, which will lead to significant economic returns of about 11.1 trillion United States (US) dollars each year \cite{Ikpehai2019LPWAN}.
\begin{figure*}
	\vspace*{-7mm}
	\centering
	\subfloat[]{\includegraphics[width=0.5\textwidth]{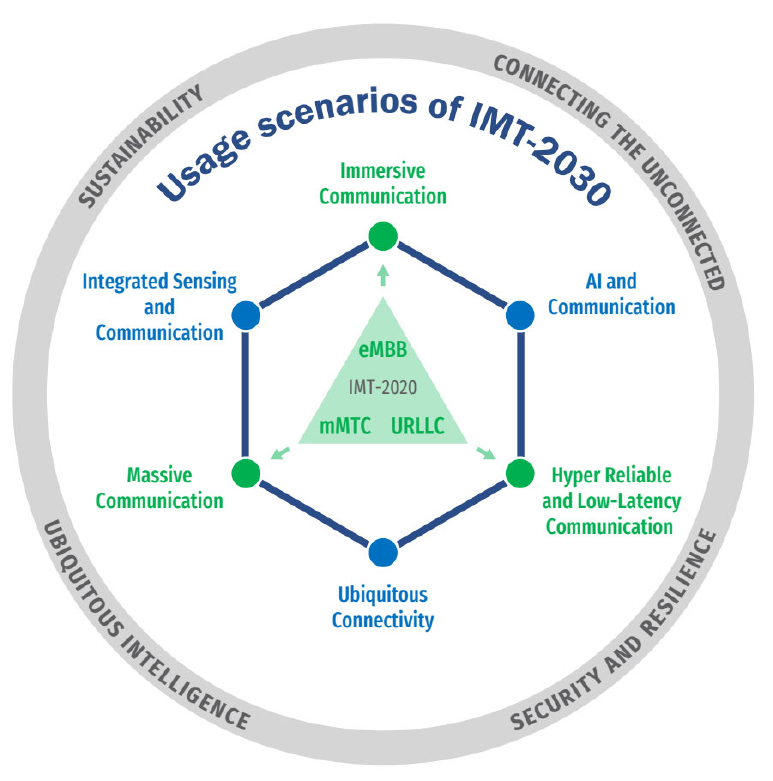}\label{Fig0a}}
	\subfloat[]{\includegraphics[width=0.5\textwidth]{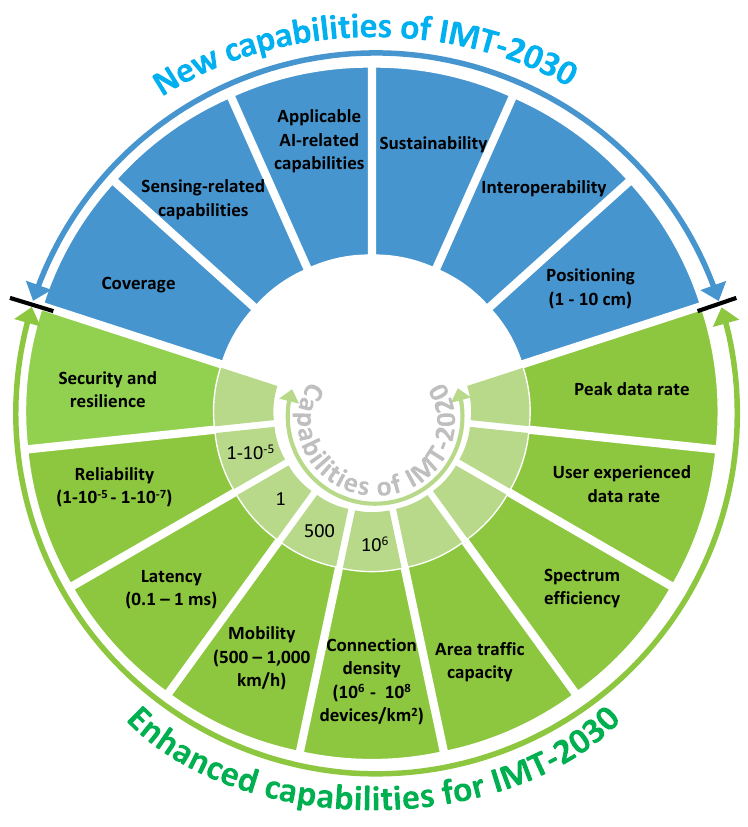}\label{Fig0b}}
	\hfill	
	\caption{(a) The usage scenarios; (b) the capabilities of 6G vision \cite{ITUR_WP_5D}.}
	\label{Fig0}
\end{figure*}

The success of IoHMT ecosystems relies on {\textit{massive communication}, which is one of the novel usage scenarios of the 6G vision, as shown in Fig.~\ref{Fig0}\,(a). Massive communication is evolved from the massive machine-type communications (mMTC) of the fifth-generation (5G) network, allowing ultra-massive numbers of machine-type devices to exchange their information either with central/distributed servers or with other devices.} Due to the highly heterogeneous nature of IoHMT applications, massive communication can be rather different from the conventional human-centric communications that are well supported in the fourth-generation (4G) and 5G cellular networks \cite{Dawy2017mMTC, Bockelmann2018mMTC, Liu2018mMTC}. Indeed, human-centric communication has the following characteristics: (i)~the downlink is usually more heavily loaded than the uplink in the support of data-hungry services provided by the core network; (ii)~the number of accommodated devices tends to be small, where the devices are relatively homogeneous such as intensively data-oriented smartphones and tablets, and their energy storage is relatively abundant due to the availability of frequent charging; (iii)~the delay requirements of various application scenarios are less stringent, e.g., the typical real-time transmissions require roughly 10 ms user-plane latency; and (iv)~the signaling overhead incurred by requesting access is not the main obstacle even for high-mobility scenarios. By contrast, mMTC exhibits the following features: (i)~the uplink typically generates dominant data traffic and its performance becomes the main bottleneck due to the exceedingly large amount of signaling overhead for massive access, while it is necessary for devices to establish connectivity with the base station (BS) via the uplink access before initiating their downlink traffic; (ii)~heterogeneous devices exhibit periodic, continuous, or event-triggered uplink traffic, and the number of simultaneously served devices can be massive; and (iii)~the associated diverse delay requirements ranging from time-critical use cases (less than 1 ms) to delay-tolerant applications (up to 100 ms) are common. {On the other hand, more stringent requirements on the capabilities have been put forward for the 6G vision, including connection density of 10\textsuperscript{6} $\sim$ 10\textsuperscript{8} device/km\textsuperscript{2}, latency of 0.1 $\sim$ 1ms, reliability of 1-10\textsuperscript{-5} $\sim$ 1-10\textsuperscript{-7}, and so on, as illustrated by Fig.~\ref{Fig0}\,(b). Clearly, there still exists a huge performance gap to bridge even with the state-of-the-art technologies.} As a result, it is urgently desired to design a new random access paradigm to embrace the IoHMT era, since the existing ones designed for human-centric communications do not facilitate the long-term evolution of machine-type communications \cite{Bockelmann2018mMTC}.

\begin{figure*}[!t]
\begin{center}
\includegraphics[width=15cm, keepaspectratio]{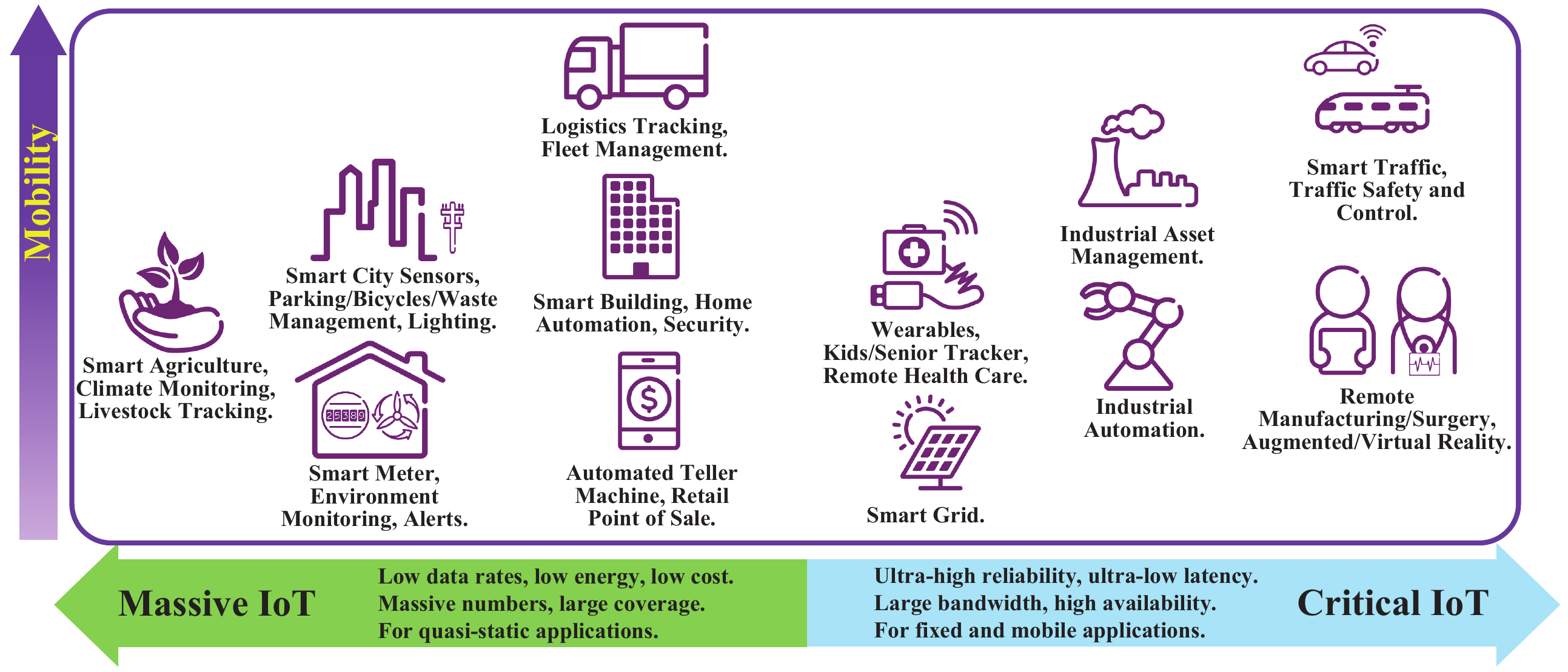}
\end{center}
\caption{\small{The application cases of future IoT can be divided into the families of massive IoT and critical IoT.}}     
\label{Fig1}
\end{figure*}

To elaborate a little further, anticipated future IoHMT applications can be classified into the families of massive IoT{\footnote{In accordance with existing standards and academic research, the term ``IoT" will be consistently used in the text that follows. This terminology encompasses both the existing IoT and the potential IoHMT in the future.}} and critical IoT according to their different service requirements, as illustrated in Fig.~\ref{Fig1}. The massive IoT family generally involves a massive number of low-cost and energy-constrained devices, supporting uplink-dominated low-data rate transmissions. Its typical applications include smart wireless sensors for monitoring, alerts, and tracking in the fields of agriculture, city management, building automation, logistics, etc \cite{Guo2021IoT}. By contrast, the critical IoT family requires ultra-high reliability and ultra-low latency for both fixed and mobile IoT scenarios. The typical applications encompass remote manufacturing/training/surgery, intelligent transport systems, smart grid, industrial automation, wearable devices, etc \cite{Khan2022IoT}. Therefore, the two families have significantly diverse service requirements, which can be summarized from various key application cases as follows.

\begin{itemize}
\item{{\bf Smart sensing, metering, and monitoring} require ultra-high density device deployments with fixed locations in a large coverage, where the devices have ultra-low power consumption, cost, and complexity. Moreover, the devices generate periodic or event-triggered low-rate traffic with small payload sizes and high delay tolerance. The data traffic is uplink dominated with few downlink control signalings being required.}

\item{{\bf Building automation, logistics, and wearable IoT} require high-density device deployments supporting low mobility, where the devices have relatively low power consumption, cost, and complexity. Moreover, the devices exhibit medium-rate bi-directional traffic while having certain delay requirements for the uplink transmission. Also, mobile devices can report geo-locations to the servers for positioning.}

\item{{\bf Intelligent transport systems} require medium-density device deployments with high mobility, where the devices generate periodic or event-triggered medium-rate traffic, and also have stringent requirements on the latency and reliability for the uplink transmission and downlink control.}

\item{{\bf Remote manufacturing, training, and surgery} require low-density device deployments with known hot spot locations, where the devices generate continuous or event-triggered high-rate traffic and also demand stringent requirements on both the latency and reliability for bi-directional payload and signaling communications.}
\end{itemize}
Clearly, one of the salient challenges of massive communication lies in designing a more efficient random access paradigm, which is expected to accommodate massive numbers of devices, reduce access latency, improve detection reliability, and satisfy heterogeneous service requirements. Generally speaking, there is a non-trivial trade-off between latency and reliability in critical IoT applications \cite{Liu2018mMTC, Bockelmann2018mMTC}. Most existing survey papers on massive access for mMTC only review the most recent advances from the academic community, while the overview of IoT standards in the industry community is limited \cite{Dai2018NOMA, Chen2020MA5G, Shahab2020NOMA}. Although different grant-based/grant-free non-orthogonal multiple access (NOMA) schemes have been reviewed, a comprehensive overview on the more specific compressive sensing (CS)-based grant-free massive access (GFMA) is still absent. This paper seeks to fullfill this gap.

In this paper, we first review the state-of-the-art IoT standards and mMTC solutions in the industry community and specify the major limitations of the existing random access schemes. Then, a comprehensive CS-based GFMA roadmap is presented, where the evolutions from single-antenna to large-scale antenna array-based base stations (BSs), from single-station to cooperative massive multiple-input multiple-output (MIMO) systems, and from unsourced to sourced random access scenarios are detailed. Finally, the key challenges and open issues are summarized. For the convenience of readers, a list of major abbreviations used in this paper is provided in the Appendix.

\vspace*{5mm}
\section{Overview of State-of-the-Art IoT Standards}\label{S2}

\begin{figure*}[!th]
\begin{center}
\includegraphics[width = 15cm]{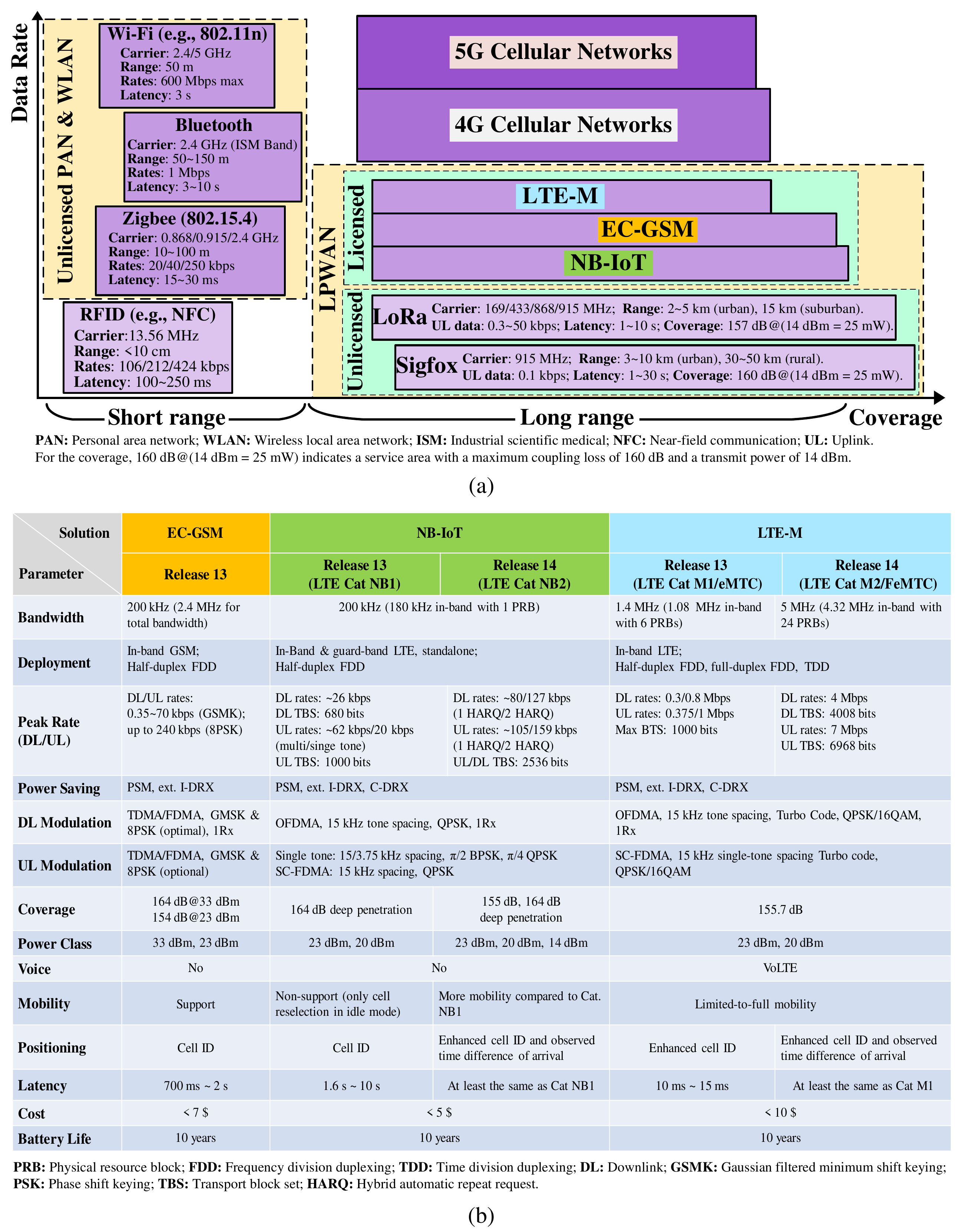}
\end{center}
\caption{\small{(a)~The whole landscape of existing IoT standards and their key features, and (b)~Massive/critical IoT connectivity can be achieved by cellular LPWANs, including EC-GSM, LTE-M, and NB-IoT standardized by the 3GPP, where the corresponding features are listed \cite{Ikpehai2019LPWAN, Chen2020MA5G, Shahab2020NOMA}.}}
\label{Fig2}  
\end{figure*}

Seamless and stable wireless connectivity is a fundamental prerequisite for designing IoT ecosystems. Owing to their heterogeneous service requirements and limited physical resources, there does not exist a single connectivity solution that can fit all emerging IoT applications. Currently, a proportion of IoT devices have been interconnected via low-cost commercial technologies, such as radio-frequency identification (RFID) \cite{Subrahmannian2022RFID}, {Bluetooth} \cite{Chang2014Bouletooth}, Zigbee \cite{Varghese2019Zigbee}, wireless fidelity (Wi-Fi) \cite{Pokhrel2018WIFI}, etc. However, these technologies only support short-range wireless communications, i.e., up to hundreds of meters, which severely hinders their practical implementations in future IoT applications that require ubiquitous coverage for widely distributed devices \cite{Chen2020MA5G, Shahab2020NOMA}. Indeed, a significant number of IoT devices will have to be connected by low-power wide-area networks (LPWANs) for better coverage. To this end, the wireless communication industry has been standardizing several LPWAN solutions, which can be divided into unlicensed and licensed LPWANs, respectively \cite{Chen2022LPWAN}. In particular, the former category, also known as (a.k.a.) non-cellular LPWAN, includes Sigfox and long range radio (LoRa), while the latter category, a.k.a. cellular LPWAN, includes extended coverage global system for mobile communications (EC-GSM), long term evolution for machine (LTE-M), and narrow-band IoT (NB-IoT) \cite{Raza2017LPWAN}. This section reports the whole landscape of existing IoT standards, in terms of data rate and coverage, and characterizes their key performance indicators, as illustrated in Fig.~\ref{Fig2}\,(a).

In practice, non-cellular LPWANs are the emerging proprietary wireless connectivity solutions designed for low-cost devices in massive IoT \cite{Ikpehai2019LPWAN}. Their advantages and drivers are low complexity and low cost, but at the expense of much lower throughput, higher latency, and susceptibility to the interference in unlicensed bands. In general, these low-cost options are still attractive to numerous enterprises interested in cheap IoT deployment. Particularly, Sigfox relies on a unified network that has been globally deployed and operated by the owing company for covering more than 60 countries and regions. Yet, the used chipset is open source since the company freely provides the protocol specifications to chip manufacturers as long as certain business terms are agreed upon \cite{Lavric2019Sigfox}. By contrast, LoRa allows the customers to flexibly establish their private networks, but the involved physical layer techniques of the chipset are proprietary to the US corporation Semtech \cite{Sundaram2020LoRa}.

On the other hand, cellular LPWANs, as listed in Fig.~\ref{Fig2}\,(b), are standardized by the third-generation partnership project (3GPP) exploiting licensed bands. Different cellular LPWANs complement each other in terms of technology availability, service requirements, and practical deployment conditions. Here, we summarize the standards of cellular LPWANs as follows.

\begin{itemize}
\item{\bf EC-GSM} was introduced in 3GPP Release 13 by adding new control and data channels to the conventional GSM networks, which can be readily achieved by applying a simple software update to the existing GSM systems \cite{Lippuner2018EC-GSM}. Note that EC-GSM still dominates many mobile markets and has been supporting a majority of cellular IoT applications via general packet radio services (GPRS). As a benefit of its backward compatibility, deploying EC-GSM based on the global coverage of traditional GSM networks can result in an extensive coverage from day one, expediting its market penetration. In general, EC-GSM has a higher uplink capacity and wider downlink coverage than legacy GSM systems, at the expense of its higher power consumption and higher complexity at the devices.

\item{\bf NB-IoT} is a clean-slate solution specifically tailored for massive low-throughput, low-cost, and energy-constrained IoT devices \cite{Ratasuk2016NB-IoT, Adhikary2016NB-IoT}. It has engaged a new power-saving mode (PSM) and the extended discontinuous reception (eDRX) for prolonging the battery life of IoT devices to 10 years or more, and achieves an extra 20 dB of power boost over the legacy GPRS \cite{Peisa2020R16&17}. NB-IoT has the advantages of reduced cost and improved energy efficiency over LTE-M. It also outperforms both Sigfox and LoRa in terms of throughput, response speed, and quality-of-services (QoS). It can be deployed either exploiting the guard-band, or within the existing 4G LTE spectrum, or as a standalone carrier relying on the second-generation (2G) spectrum. However, the handovers among different cells would be a problem for NB-IoT, due to the high control signaling overhead, which makes it the best suited for static setting rather than mobile devices.

\item{\bf LTE-M} is the most flexible LPWAN solution supporting a full breadth of IoT application cases, varying from low-end static sensors to high-end mobile devices requiring high throughput~\cite{Ratasuk2014LTE-M}. It also has PSM and eDRX strategies that can be multiplexed onto the full LTE carriers. Hence, it is eminently suitable for co-existence with existing cellular networks \cite{Peisa2020R16&17}. The advantages of LTE-M over EC-GSM and NB-IoT include higher data rate, higher mobility, and the support of voice communications, albeit at the expense of requiring more bandwidth and a higher implementation cost. 
\end{itemize}

Note that both NB-IoT and LTE-M were introduced in Release 13 and have evolved to Release~17 at the time of writing. Fig. \ref{Fig2}\,(b) compares EC-GSM, NB-IoT (Releases 13 $\&$ 14), and LTE-M (Releases 13 $\&$ 14), and the readers can refer to \cite{Lippuner2018EC-GSM, Ratasuk2016NB-IoT, Adhikary2016NB-IoT, Peisa2020R16&17, Ratasuk2014LTE-M} for their detailed technical parameters, including bandwidth, peak rate, modulation type, latency, cost, battery life, etc. Compared to Releases 13 $\&$ 14, Releases 15 $\&$ 16 for NB-IoT and LTE-M have further improved the spectral and energy efficiencies. Moreover, Release 17 has carried out a study on the possibility and required specification updates to support NB-IoT and LTE-M in non-terrestrial networks. Considering the random access of NB-IoT and LTE-M, Release 15 introduced an early data transmission (EDT) mode, while Release 16 further enhanced the uplink data payload of EDT and introduced a pre-configured uplink resources (PUR) mode \cite{Peisa2020R16&17}. These emerging strategies can reduce the overhead for signaling exchanges, thus improving the system energy efficiency and reducing the random access latency, which will be detailed in the next section.

\section{Existing Random Access Solutions and Limitations}\label{S3}

Efficient random access protocols and multiple access techniques are the fundamental premises for connecting a massive number of devices. Compared to the classical grant-based four-step random access (FSRA) developed in 4G LTE \cite{4S-GBMA}, the industry community has gradually simplified the random access procedure in recent 3GPP releases. Due to the need for the contention or the pre-configuration of orthogonal physical resources for avoiding inter-device interferences, these solutions belong to the grant-based/grant-free orthogonal multiple access (OMA) paradigms. However, the maximum number of accommodated devices is limited by the number of orthogonal resources. To overcome this limitation, various non-orthogonal multiple access (NOMA) solutions have also been intensively investigated in the academic community. Nevertheless, most of them have inherent limitations in supporting future massive IoT connectivity.

\subsection{Random Access in Industry Standardization}\label{S3.1}

\begin{figure*}[!t]
\begin{center}
\includegraphics[width = 15cm]{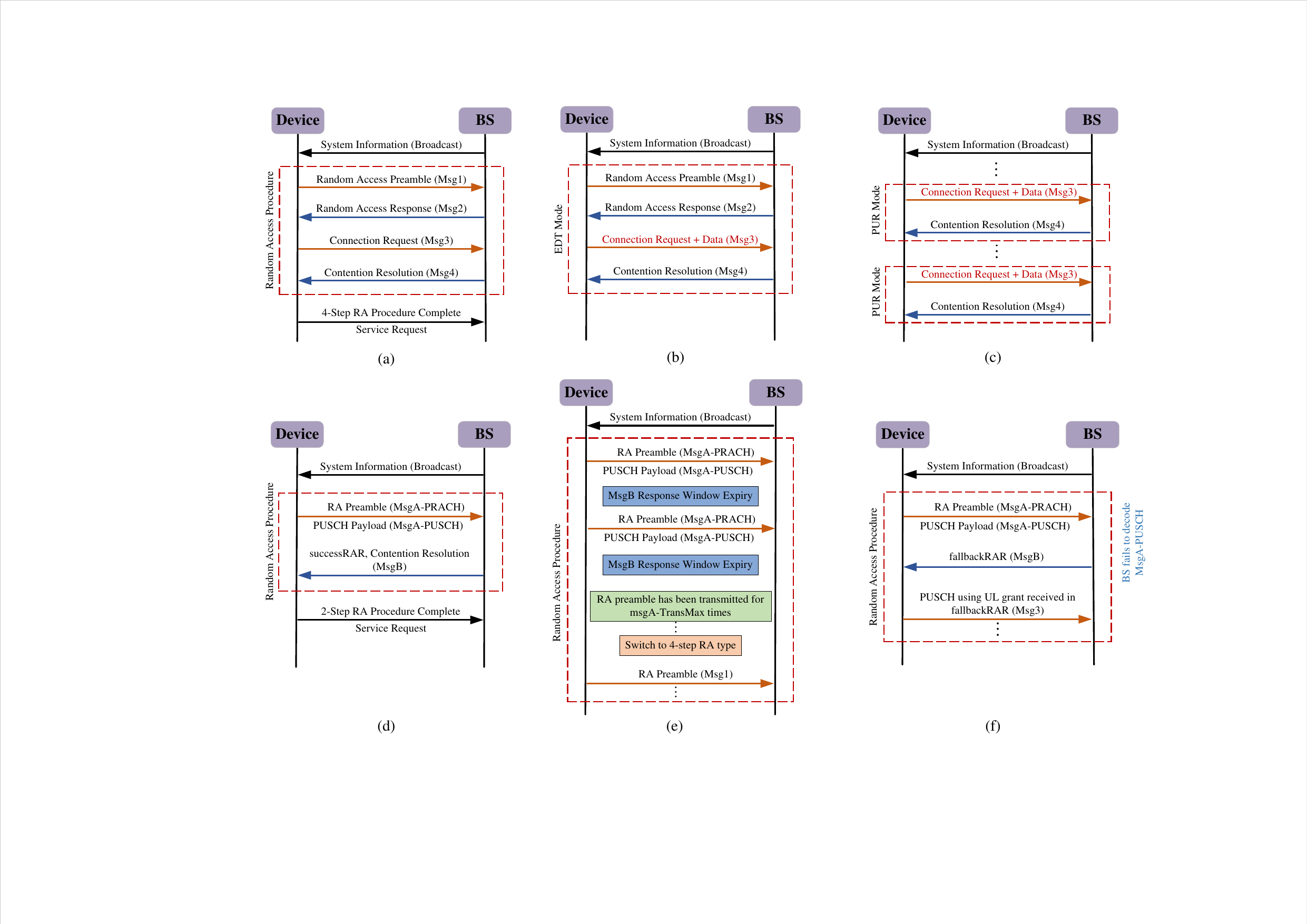}
\end{center}
\caption{\small{(a)~Standard FSRA in 4G LTE, (b)~FSRA with EDT mode, (c)~PUR-based random access, and (d)-(f)~respectively correspond to three cases of TSRA in 5G NR.}}     
\label{Fig3}
\end{figure*}

In unlicensed bands, both Sigfox and LoRa adopt ALOHA-based random access schemes, where the devices exploit a random frequency and time division multiple access technique to transmit their signals \cite{Ferre2018ULPWAN}. Without the need for direct signaling interactions between the devices and the BS to establish connection, these schemes can be classified into the grant-free OMA category. Although the related access procedure is simple enough, the severe collision is a limiting factor for the realization of low-latency and high-efficiency random access when the number of devices becomes large. In licensed bands, similar to 4G LTE, the early NB-IoT and LTE-M in Release 13 adopt the grant-based FSRA protocol illustrated in Fig.~\ref{Fig3}\,(a), which includes Msg1-Msg4 \cite{Shahab2020NOMA}. More specifically, the FSRA procedure is summarized as follows.

\begin{itemize}
\item{{\bf Step 1:} According to the system information periodically broadcasted by the BS, the requesting devices transmit a contention-based orthogonal preamble, i.e., Msg1, on the uplink physical random access channel (PRACH).}
	
\item{{\bf Step 2:} After successfully receiving the preamble, the BS broadcasts a random access response (RAR), i.e., Msg2, which encompasses the detected preamble identification, time-alignment instructions for uplink synchronization, temporary cell radio network temporary identifier (TC-RNTI), and the uplink grant for Msg3.} 
	
\item{{\bf Step 3:} Upon receiving the RAR within a given time window, the device, whose Msg1 is successfully detected by the BS, transmits its connection request, i.e., Msg3, via a physical uplink shared channel (PUSCH) indicated in Msg2. If the RAR is not received within the given time window, this round random access predicates failure.}
	           
\item{{\bf Step 4:} After receiving Msg3, the BS replies to the devices with a contention resolution message, i.e., Msg4. If a device finds its contention resolution identification in Msg4, an acknowledgment is fed back to the BS, the FSRA procedure is completed and the granted device moves to the connected mode. Otherwise, a new access schedule is attempted.}
\end{itemize}

The grant-based FSRA procedure necessitates two round-trip interactions between the devices and the BS to establish connection, which results in significant signaling overhead and long access latency. Moreover, the number of available orthogonal preambles is generally limited due to the limited number of physical resources. Typically, a fraction of the total 64 orthogonal preambles are reserved for contention-free access (e.g., handover) and only the remaining orthogonal preambles can be adopted for contention-based random access \cite{Peisa2020R16&17}. Therefore, as the number of simultaneously served devices becomes large, the access efficiency is significantly degraded due to high collision probability. To improve the access efficiency for NB-IoT and LTE-M, the EDT and PUR modes were introduced in Release 15 and Release 16, respectively \cite{Hoglund2020R16}. 

As depicted in Fig.~\ref{Fig3}\,(b), for the EDT mode, a device in the idle state may directly transmit a maximum of 1000 data bits embedded in Msg3 of FSRA. After a successful reception at the BS, the device may remain in the idle state or switch to the connected state for further data transmission. In other words, if a device has short data ($\le 1000$\,bits), the data can be transmitted in a single successful FSRA without the need to go to the connected mode. On the other hand, for long data transmission, a device may rely on the grant-based FSRA to establish the connection.

By contrast, PUR is tailored for quasi-static devices conveying periodic or pseudo-varying short data traffic. Specifically, the devices adopting the PUR mode are pre-configured with uplink transmission resources via dedicated radio resource control signaling. Thus, Msg1 and Msg2 are omitted, and data can be directly delivered in Msg3, as illustrated in Fig.~\ref{Fig3}\,(c). In practice, the PUR mode has two categories: dedicated PUR and shared PUR. The former is designed for devices conveying periodic traffic, where the uplink time-frequency resources are exclusive for each device, while for the latter, the same uplink time-frequency resources are shared by up to two devices, where the superimposed signals can be distinguished by mutually orthogonal demodulation reference signals (DMRS). Clearly, the PUR-based random access is a grant-free OMA solution, which cannot accommodate a large number of devices due to the limited number of orthogonal resources and DMRS.

In contrast to FSRA, the recent Release 16 introduces a grant-based two-step random access (TSRA) protocol for the 5G new radio (NR) \cite{Hoglund2020R16}. By resorting to a simplified single round-trip interaction between the devices and the BS, the access latency and control signaling overhead can be reduced. In particular, the TSRA combines the PRACH preamble and PUSCH payload as a single MsgA transmitted by the devices, and then merges RAR and contention resolution message into a single MsgB. Figs.~\ref{Fig3}\,(d)$\sim$(f) portray three typical cases of the TSRA. Specifically, 

\begin{itemize}
\item{{\bf Case 1:} Upon the device receiving a MsgB with successful RAR and contention resolution, the grant-based TSRA is completed and the device moves to the connected mode, as illustrated in Fig.~\ref{Fig3}\,(d)}.

\item{{\bf Case 2:} If the device fails to receive a MsgB after a maximum number of MsgA trials, the device switches to the grant-based FSRA mode by transmitting Msg1 for a new access attempt, as shown in Fig.~\ref{Fig3}\,(e)}.

\item{{\bf Case 3:} If the MsgA cannot be decoded correctly, the BS broadcasts a MsgB with fallback RAR, and the device falls back to the grant-based FSRA mode by transmitting Msg3 for connection request, see Fig.~\ref{Fig3}\,(f).}
\end{itemize}

Note that only 5G NR supports the grant-based TSRA, while NB-IoT and LTE-M currently do not have this mode. At the time of writing, all the standardized random access solutions for cellular LPWANs adopt OMA techniques, i.e., exploiting orthogonal radio resources to distinguish different devices, and thus they belong to the grant-based/grant-free OMA paradigms. Here, almost all these solutions require a grant-based random access procedure for orthogonal resource contention. The only exception is the PUR-based random access, where the orthogonal resources are pre-configured. Specifically, for Msg1 or MsgA, the PRACH preamble of each active device is anonymously selected from a predefined orthogonal sequence pool, while for shared PUR, two devices are distinguished via orthogonal DMRS. 

However, the number of orthogonal radio resources is limited but the number of requesting devices has been exponentially increasing, since the rise of the IoT. Therefore, all these OMA solutions suffer from unavoidable preamble collision with degraded access efficiency, which eventually leads to the network congestion \cite{Shahab2020NOMA, Peisa2020R16&17, Hoglund2020R16}.

\subsection{Review of Non-Standardized NOMA}\label{S3.2}

Compared with OMA, NOMA has the potential to accommodate a larger number of devices, which can be simultaneously served over a small amount of time-frequency resource elements by exploiting the devices' unique but non-orthogonal signatures. A NOMA scheme is generally a grant-free solution, offering advantage of reducing access signaling overhead and latency. There are potentially many NOMA techniques, but the most popular ones are classified into two categories: power-domain NOMA and code-domain NOMA \cite{Dai2018NOMA}. More recently, spatial-domain NOMA has attracted much attention, which relies on MIMO technology to support multiple users on the same time-frequency resources via uplink multiuser detection (MUD) and downlink multiuser transmit (MUT) precoding/beamforming \cite{SD-NOMAul2008,Zhang2011SDMA-2,Zhang2014SDMA,SD-NOMAul2020,SD-NOMAdl2009,SD-NOMAdl2016,SD-NOMAdl2020,SD-NOMAdl2021}.

\subsubsection{Power-domain NOMA}\label{S3.2.1}

The signals of multiple devices are superimposed on the same time-frequency resources with different power levels, which requires dedicated power allocation strategies at the transmitter and the successive interference cancellation (SIC)-based data detection at the receiver for efficient decoding \cite{Benjebbour2013NOMA, Saito2013NOMA, Higuchi2013NOMA, Nonaka2014NOMA}. However, power-domain NOMA is usually limited to serving a small number of human-type devices. Specifically, it is challenging to ensure distinguishable power levels for massive devices, particularly in grant-free random access. Furthermore, it is also impractical to design a reliable SIC-based receiver, since the power levels are generally not distinctive while the resolutions of analog-to-digital converters (ADCs) are limited. In practice, the users have to be divided into groups, each containing only a small number of users. The users in the same group can adopt a power-domain NOMA scheme for transmission, while the users in different groups have to employ OMA schemes for avoiding inter-group interferences. Therefore, a grant-based scheduling procedure is required, which will result in extra signaling overhead and latency.

\subsubsection{Code-domain NOMA}\label{S3.2.2}

On the other hand, code-domain NOMA realizes multiplexing in the code domain, where multiple devices share the same time-frequency resources but adopt non-orthogonal low-cross-correlation spreading sequences as their unique signatures \cite{Hoshyar2008NOMA, Guo2008NOMA, Beek2009NOMA, Razavi2011NOMA, Lu2015SCMA, Yuan2016MUSA}. The superimposed signals of different devices are distinguished by leveraging the uniqueness of spreading sequences. This category of NOMA schemes is inspired by the classical code-domain multiple access (CDMA) utilizing orthogonal spreading sequences to distinguish devices. The key difference lies in that the spreading sequences of code-domain NOMA are non-orthogonal low-cross-correlation sequences, and thus more devices can be accommodated given the same amount of physical resources. Compared with power-domain NOMA, code-domain NOMA is superior in supporting massive IoT connectivity with grant-free random access, since the number of available spreading sequences is much larger than the resolution of the received power levels. In general, various code-domain NOMA schemes can be further categorized into low-density spreading (LDS) NOMA class and dense spreading (DS) NOMA class, depending on whether the adopted spreading sequences are sparse or not.

In practice, LDS NOMA schemes, such as the intensively investigated LDS-CDMA, LDS-OFDM, and sparse code multiple access (SCMA), adopt sparse spreading codes, which can be transmitted either in the time or frequency domain. Compared with the dense spreading codes of conventional CDMA, the sparse codes of LDS NOMA can still offer certain spreading gains to suppress the undesired inter-device interferences, while facilitating the application of low-complexity message passing algorithms for data detection \cite{Wei2019SCMA}. LDS-CDMA and LDS-OFDM are two initial code-domain NOMA schemes that are directly extended from the traditional CDMA and OFDM cellular systems. In particular, for LDS-CDMA, the symbol to be transmitted is spread over the time domain, while for LDS-OFDM, the chips are transmitted in the frequency domain. Developed from basic LDS-CDMA, the state-of-the-art SCMA scheme directly maps the transmitted bit streams onto a set of sparse codewords, which results in shaping gains, leading to an improved detection performance. Specifically, let us consider the SCMA schematic diagram of Fig. 3 in~\cite{Gao2018CS} as an example, where each device has a unique sparse codebook of $L_{\rm SCMA} = 4$ sparse codewords such that $(\log_2{L_{\rm SCMA}})$-bit information per channel use can be delivered. In this context, $K_{\rm SCMA} = 6$ devices share $N_{\rm SCMA}=4$ resource elements and this is termed as a subcarrier block, where the overloading rate is defined as $\gamma = K_{\rm SCMA}/N_{\rm SCMA} = 150\%$. Moreover, the $K$ devices' sparse codebooks can be reused in multiple subcarrier blocks for joint transmission and decoding. Note that sparse codes of the same device share the same sparse pattern (i.e., the positions of non-zero elements in a vector), which is unique for each device.

On the other hand, multi-user shared access (MUSA) is a typical DS NOMA scheme, where a set of non-orthogonal dense spreading sequences constitutes a pool and each device anonymously chooses a sequence to spread its transmit symbol \cite{Yuan2016MUSA}. Compared with the LDS-CDMA and LDS-OFDM where multiple resource blocks reuse the same LDS sequence, the devices in MUSA may pick different sequences for spreading different symbols, attaining an enhanced performance with the aid of interference averaging. Another difference between the two classes lies in their spreading sequence configuration mode. Devices in LDS-CDMA and LDS-OFDM are pre-configured to employ unique spreading sequence, while MUSA adopts the so-called contention-based random access, where devices share the same pool of sequences. More details on SCMA and MUSA can be found in~\cite{Lu2015SCMA, Yuan2016MUSA, Wei2019SCMA}.

Now Let us consider the sporadic traffic generated by massive communication. Although the number $K$ of devices to be served can be hundreds even thousands, the number of simultaneously active devices $K_a$ is relatively small \cite{Liu2018mMTC}. For example, $K = 480$ and $K_a = 48$ result in an activity ratio of $\rho = 0.1$. Unfortunately, most existing NOMA schemes are incapable to cope with this sparse traffic. Taking SCMA as an example, designing high-dimensional sparse codebooks for simultaneously supporting devices for a large $K$ is quite challenging, since the sparse teletraffic of massive communication cannot be readily exploited. Considering the sparse codebooks designed in~\cite{Wei2019SCMA} for $K_{\rm SCMA}=48$ and $N_{\rm SCMA}=24$, SCMA has to divide $K=480$ total devices into 10 SCMA groups having a total of $10N_{\rm SCMA}=240$ resource elements. The devices in the same group adopt SCMA scheme for transmission, while the devices in different groups employ OMA schemes for avoiding inter-group interferences. In this context, a grant-based scheduling procedure may be required, which results in extra signaling overhead and latency.  

\subsubsection{Spatial-domain NOMA}\label{S3.2.3}

Spatial-domain NOMA, also known as space division multiple access in some earlier literature \cite{Vand2001SDMA}, exploits the extra spatial degrees-of-freedom (DoF) offered by MIMO techniques to realize multiplexing. Specifically, the signals of multiple devices conveyed on the same time-frequency resources are distinguished by exploiting unique user-specific channel impulse responses (CIRs). Intuitively, the user-specific CIR plays a role similar to the non-orthogonal spreading sequence in code-domain NOMA. Given the users' CIRs, therefore, the signals of multiple devices conveyed on the same time-frequency resources can be recovered using MUD at the uplink BS receiver \cite{SD-NOMAul2008,Zhang2011SDMA-2,Zhang2014SDMA,SD-NOMAul2020} or distinguished through MUT precoding at the downlink BS transmitter \cite{SD-NOMAdl2009,SD-NOMAdl2016,SD-NOMAdl2020,SD-NOMAdl2021}. Hence spatial-domain NOMA requires accurate estimate of users' CIRs or MIMO channel state information (CSI). Acquisition of accurate MIMO CSI however imposes considerable pilot resource overhead, which may be unaffordable in practice. In order to attain near-optimal performance based on the limited pilot resources, joint channel estimation and turbo MUD/decoding solutions have attracted substantial research interests \cite{Zhang2011SDMA-2, Zhang2014SDMA}. However, these joint channel estimation and data detection solutions are generally computationally expensive, and they can only support limited number of users. 

By deploying large-scale or massive antenna array at BS, favourable massive MIMO (mMIMO) environment is created for implementing spatial-domain NOMA. In particular, the CIRs associated with different users become nearly orthogonal, and the signals of multiple devices can be separated with low-complexity conjugate beamforming \cite{Larsson_etal2014}. But acquisition of the mMIMO CSI becomes even more challenging. In order to achieve affordable-complexity mMIMO CSI estimation, orthogonal pilot sequences must be adopted. However, the
number of orthogonal pilot sequences available is limited, and these pilot resources will have to be reused in neighbouring cells. This causes severe pilot contamination which results in the BS being unable to reliably differentiate the signals of different cells. Sophisticated pilot designs \cite{PCE2013,PCE2014,PCE2016-Aug,PCE2016-Nov} have been developed to mitigate pilot contamination. In the past decade, the emerging of mMIMO techniques has accelerated the development of spatial-domain NOMA. However, the number of devices that can be supported by the existing spatial-domain NOMA solutions is still limited, and it is very challenging to apply these existing spatial-domain NOMA techniques to support a massive number of devices in future massive communication scenarios.

\section{Compressive Sensing-Based Grant-Free Massive Access Paradigm} \label{S4}

The discussions in the previous sections reveal that neither the current standardized OMA solutions nor the existing non-standardized NOMA solutions can well accommodate future IoT applications with massive communication. To tackle this issue, a CS-based GFMA paradigm was recently developed, where the active devices directly transmit their uplink access signals over the same time-frequency resources without the need for any scheduling in advance \cite{Bayesteh2014GFMA, Zhang2016GFMA}. Meanwhile, by leveraging the sporadic traffic of devices, the multi-device detection at the BS can be formulated as a CS problem that can be effectively resolved by various sparse signal recovery algorithms \cite{Senel2018GFMA}. Therefore, the complicated signaling interactions for access scheduling, including resource granting and contention resolution, associated with grant-based random access protocols are circumvented. Furthermore, compared with conventional NOMA schemes, the designs of distinguishable access signatures and the multi-device detection algorithm are significantly simplified by further taking into account the sporadic traffic.

To begin with, the essence of the CS theory can be well captured in the following discussion starting with the mathematical expression of
\begin{equation}\label{Eq1}
 {\bf Y} = {\bf \Phi} {\bf X} + {\bf N},
\end{equation}
where ${\bf Y}\in \mathbb{C}^{M \times Q}$ represents the low-dimensional measurements, ${\bf \Phi} \in \mathbb{C}^{M \times N}$ is the sensing matrix with $M \ll N$, ${\bf X} \in \mathbb{C}^{N \times Q}$ is the original high-dimensional sparse signal, and $\bf N$ is the additive white Gaussian noise (AWGN). The CS theory indicates that given ${\bf Y}$ and ${\bf \Phi}$, the sparse matrix ${\bf X}$ can be exactly recovered as long as $M~\ge~N_a{\rm log}_2{\left(N\right)}$ is satisfied \cite{Gao2018CS}. Here, $N_a \ll N$ is the maximum number of non-zero elements in the columns of ${\bf X}$. As such, the classic model in (\ref{Eq1}) becomes a standard single-measurement vector (SMV) CS problem for $Q = 1$ \cite{Candes2008CS} and a multiple-measurement vector (MMV) CS problem for $Q > 1$ \cite{Chen2006MMV, Davies2012MMV, Cotter2005MMV}.

The key challenge in solving the CS recovery problem is how to design a computationally efficient sparse signal recovery algorithm. Various CS recovery algorithms have been proposed, which can be classified into three categories, namely, convex relaxation algorithms, greedy-based algorithms, and Bayesian inference algorithms. Specifically, convex relaxation algorithms, such as basis pursuit \cite{Candes2006BP} and least absolute shrinkage and selection operator (LASSO) \cite{Tibshirani1996LASSO}, relax the non-convex CS recovery problem as a conventional convex optimization problem and employ linear programming methods to acquire the solution. These algorithms generally enjoy an excellent recovery performance but the related computational complexity is extremely high, especially for large problem dimensions. By contrast, greedy-based algorithms, such as orthogonal matching pursuit (OMP) \cite{Pati1993OMP}, subspace pursuit (SP) \cite{ Dai2009SP}, and CoSaMP \cite{Needel2009CoSaMP}, identify the non-zero indices of ${\bf X}$ and estimate their corresponding coefficients in a greedy iterative manner. In general, they have low algorithmic complexity but suffer from significant performance losses when the number of measurements is relatively small or the signal-to-noise ratio (SNR) is low. Both convex relaxation and greedy-based algorithms only consider the sparsity of ${\bf X}$ but fail to leverage its statistical information for effectively improving recovery accuracy. To overcome this limitation, Bayesian inference algorithms, such as belief propagation \cite{Kschischang2001BP}, expectation propagation (EP)~\cite{Minka2001EP}, and sparse Bayesian learning (SBL) \cite{Tipping2001SBL}, were developed under the Bayesian framework, by establishing various flexible {\em a priori} distributions to capture the sparsity properties and the statistical information of ${\bf X}$, thus reaping a better recovery performance. Moreover, the {trade off} between recovery performance and computational complexity can be effectively achieved by a low-complexity approximation to the standard Bayesian inference framework, known as the approximate message passing (AMP) framework \cite{Dohono2010AMP}.

\begin{figure}[!t]
\begin{center}
\includegraphics[width=1\columnwidth, keepaspectratio]{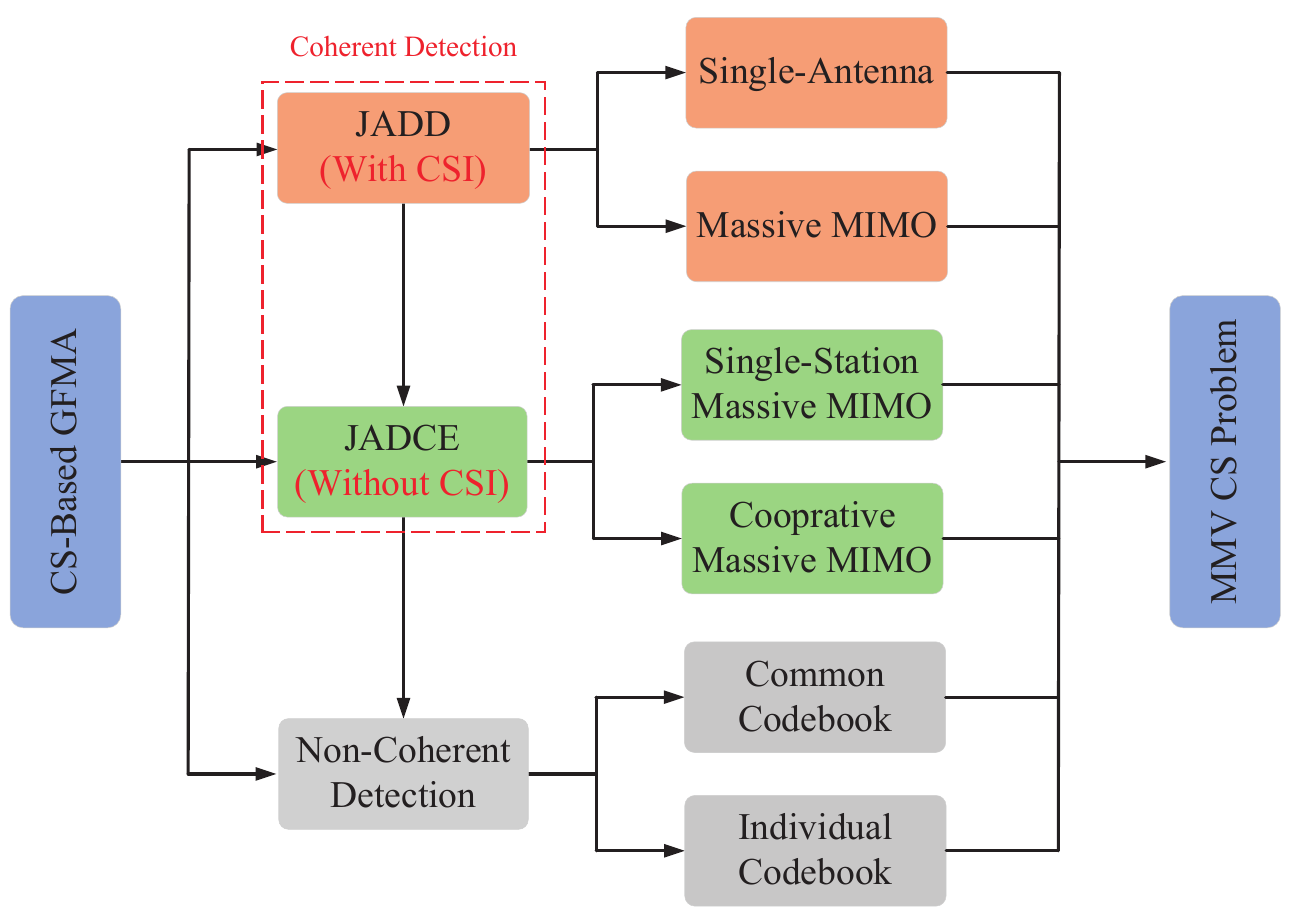}
\end{center}
\caption{\small{The classification of existing CS-based GFMA schemes. The evolutions from single-antenna to large-scale antenna array-based BSs, from single-station to cooperative massive MIMO systems, and from unsourced to sourced access scenarios are illustrated.}}     
\label{Fig4}
\end{figure}

This section presents the roadmap of the promising CS-based GFMA paradigm, as illustrated in Fig.~\ref{Fig4}, where various specific GFMA schemes are introduced to fulfill the heterogeneous massive communication requirements of practical IoT applications. In particular, the evolutions from single-antenna to large-scale antenna array-based BSs, from single-station to cooperative massive MIMO systems, and from unsourced to sourced access scenarios are detailed in this section. A dominated common characteristic of these schemes is that the multi-device detection at the BS, i.e., activity detection and channel estimation (or data detection), can be formulated as a CS problem of (\ref{Eq1}). Moreover, the sparse matrix ${\bf X}$ generally exhibits different structured sparsity properties in different access scenarios, which can be exploited to further improve recovery performance with the aid of the bespoke algorithms.

\subsection{Joint Activity and Data Detection}\label{S4.1}

For the CS-based GFMA paradigm, the wireless transceiver can be flexibly designed to accommodate the practical heterogeneous massive communication requirements, resulting in various specific GFMA schemes. Assume that the CSI is available at receiver. Note that this assumption is valid in the scenarios where the channels can be efficiently estimated with very low pilot overhead or the CSI remains unchanged for a long time, such as single-antenna systems and fixed sensor networks, respectively \cite{Wang2015JADD}. With the CSI, BS can jointly detect the active devices and their payload data from the overlapped received signal. Focusing on this joint activity and data detection (JADD) problem, this subsection first investigates CS-based GFMA schemes in single-antenna systems. Then the problem is extended to massive MIMO systems, where the additional spatial DoF is exploited to enhance uplink throughput and improve detection performance.

\begin{figure}[!t]
\begin{center}
\includegraphics[width=1\columnwidth, keepaspectratio]{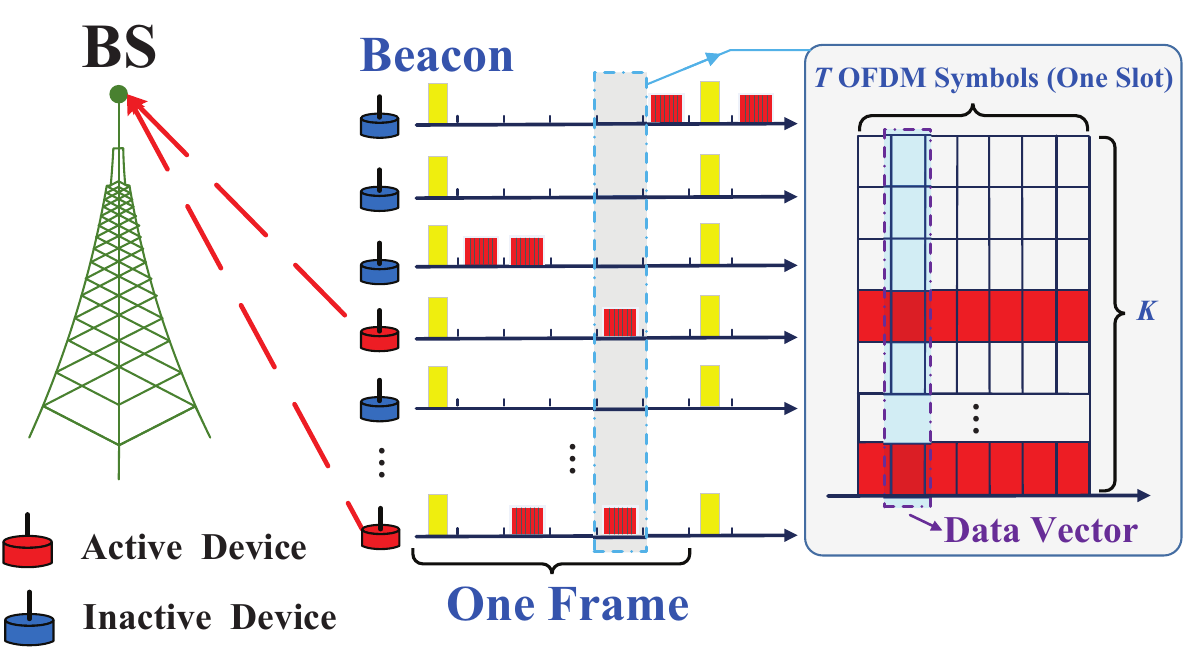}
\end{center}
\caption{\small{The system model of spreading-based GFMA scheme in single-antenna systems, where the temporal structured sparsity of the signal matrix is illustrated \cite{Mei2022JADD} \textcircled{c}IEEE.}}     
\label{Fig5}
\end{figure}

\subsubsection{JADD for GFMA in Single-Antenna Systems}\label{S4.1.1}

The GFMA in single-antenna systems generally adopts spreading-based transmission scheme. Consider a massive IoT connectivity scenario with one single-antenna BS serving $K$ single-antenna devices, where $K$ is usually large and OFDM is adopted to combat the time dispersion effect, as illustrated in Fig. \ref{Fig5}. Due to the sporadic uplink traffic, only $K_a$ ($K_a \ll K$) out of the total $K$ devices are active during each time slot with $T$ OFDM symbols, in which the activity and CSI remain unchanged. The BS periodically broadcasts its beacon signals to facilitate synchronization, power control, and channel estimation at the devices. Since only one single antenna is considered at both the devices and the BS, the beacon signal overhead is very small for downlink channel estimation. To distinguish the $k$th device from others at the BS, its access signal, $x_{k,t} \in \mathbb{C}$, in the $t$th OFDM symbol duration ($1 \le t \le T$) is spread across $L$ subcarriers by a unique spreading sequence ${\bf{s}}_k \in \mathbb{C}^{L \times 1}$. Moreover, benefiting from the channel reciprocity, the downlink CSI estimates are exploited for pre-equalizing in the uplink transmission to pre-compensate the impact of uplink channels at the devices. The active devices can directly transmit their spread access signals on the exactly same time-frequency resources, without any scheduling in advance. This avoids the complicated signaling interactions in FSRA or TSRA, and hence significantly reduces the access latency. 

Adopting the aforementioned spreading-based GFMA, the signals of all the active devices are overlapped at the BS, which results in severe inter-device interferences. Therefore, it is essential to design a reliable JADD scheme at the BS, which can be formulated as a MMV CS problem of (\ref{Eq1}). Specifically, the sensing matrix is expressed as ${\bf \Phi} = \left[{\bf s}_1, {\bf s}_2, \cdots, {\bf s}_K\right] \in \mathbb{C}^{M \times N}$ with $M = L$ and $N = K$, and the sparse signal matrix is expressed as ${\bf X} = \left[\alpha_1{\bf x}_1, \alpha_2{\bf x}_2, \cdots, \alpha_K{\bf x}_K\right]^{\rm T} \in \mathbb{C}^{N \times Q}$ with $Q = T$.
Here, the binary variable $\alpha_k \in \{ 0,1 \}$ denotes the activity indicator with 1 being active and 0 otherwise, and ${\bf x}_k \in \mathbb{C}^{T \times 1}$ is the spread access signal of the $k$th device.

Note that the device activity and payload data are embedded in the access signal matrix ${\bf X}$, i.e., the indices of non-zero rows indicates the identities of active devices and the corresponding coefficients are the transmitted signals. Hence, the JADD problem is equivalent to reconstructing ${\bf X}$ based on the overlapped received signal ${\bf Y}$ and the spreading matrix ${\bf \Phi}$. Based on the estimate of ${\bf X}$, the payload data can be further detected. To this end, the authors in \cite{Wang2015JADD} developed a CS-message passing algorithm (MPA) detector, where the CoSaMP algorithm and MPA are employed for CS-based active device detection and payload data detection, respectively, but only a single OFDM symbol ($T=1$) is considered in \cite{Wang2015JADD}. In practice, the data packet of active devices usually occupies several consecutive OFDM symbols \cite{Abebe2016JADD}. Therefore, the system is synchronized in a slot structure\footnote{In the frame structure of 5G NR, each OFDM frame consists 10 subframes with each subframe having several time slots. The number of time slots within each subframe depends on the subcarrier spacing and each slot consists of 7 OFDM symbols~\cite{3GPP2018NR}.} and the device activity remains constant during each slot, which leads to the \emph{temporal common sparsity} pattern, as illustrated in Fig. \ref{Fig5}. Furthermore, although the device activity may change across different time slots, the variation is gradual, i.e., the active devices generally transmit their data in consecutive time slots (i.e., burst transmission) with a high probability. This leads to the temporal correlation over several consecutive time slots, which is referred to as \emph{temporal dynamic sparsity}. Under this context, various JADD algorithms have been proposed to leverage these two temporal structured sparsity properties for improved detection performance. For instance, the authors in \cite{Wang2016JADD} proposed a structured iterative support detection (SISD) algorithm to leverage the temporal common sparsity, which follows the idea of greedy-based CS recovery algorithms.   By resorting to the Bayesian inference framework, a joint expectation maximization AMP (EM-AMP) algorithm was developed to further exploit the {\em a priori} statistical information of the transmitted discrete symbols \cite{Wei2017JADD}. However, both SISD algorithm and EM-AMP algorithm reconstruct the access signals of different symbol durations separately, which fails to take the full advantages of the temporal common sparsity. Hence, in \cite{Du2018JADD}, the authors proposed a block sparsity-based detection algorithm, which vectorizes ${\bf X}$ into a block-sparse vector for a better use of the temporal common sparsity based on the block CS theory \cite{Stojnic2009BCS}. In addition, an orthogonal AMP with accurate structure learning (OAMP-ASL) algorithm was also proposed in \cite{Mei2022JADD}, where the temporal common sparsity is incorporated in the {\em a priori} distribution of ${\bf X}$ for further improving performance.

\begin{table*}[!t] 
\caption{\small{Summary of JADD algorithms for spreading-based GFMA scheme in single-antenna systems}} 
\begin{center}
\resizebox{\linewidth}{!}{
\begin{tabular}{|l|c|c|l|l|l|}
\hline
\multicolumn{1}{|c|}{JADD Scheme}					&	CS Model	&	CS Algorithm			&	\multicolumn{1}{c|}{Sparsity Structure}		&	\multicolumn{1}{c|}{Advances} & {Complexity}\\
\hline
CS-MPA algorithm \cite{Wang2015JADD}				&	SMV			&	Greedy \& Bayesian		&	None										&	Exploit the sporadic uplink traffic of devices & {${\cal O}\left(KLK_a + q^w\right)$}\\ 
\hline
SISD algorithm \cite{Wang2016JADD}					&	MMV			&	Greedy					&	Temporal common sparsity					&	Exploit the temporal common sparsity & {${\cal O}\left(TK^3\right)$}\\ 
\hline
EM-AMP algorithm \cite{Wei2017JADD}					&	MMV			&	Bayesian				&	Temporal common sparsity					&	\begin{tabular}[c]{@{}l@{}} Further exploit the \emph{a priori} statistical information \\ of the transmitted discrete symbols \end{tabular} & {${\cal O}\left(TKL\right)$}\\ 
\hline
Block sparsity-based algorithm \cite{Du2018JADD}	&	SMV			&	Greedy					&	Block sparsity								&	Make better use of the temporal common sparsity & {${\cal O}\left(T^2KL+T^2Ls^2+T^3s^3\right)$} \\ 
\hline
OAMP-ASL algorithm \cite{Mei2022JADD}				&	MMV			&	Bayesian				&	Temporal common sparsity					&	\begin{tabular}[c]{@{}l@{}} Incorporate the temporal common sparsity in the \\ \emph{a priori} distribution of the access signal matrix \end{tabular} & {${\cal O}\left(TKL+T^2K+TKq\right)$}\\ 
\hline
Dynamic CS-based algorithm \cite{Wang2016DCS}		&	MMV			&	Greedy					&	Temporal dynamic sparsity					&	Exploit the temporal dynamic sparsity & {${\cal O}\left(TKLK_a + TLK_a^ 3\right)$} \\ 
\hline
PIA-ASP algorithm \cite{Du2017JADD}					&	MMV			&	Greedy					&	Temporal dynamic sparsity					&	Make better use of the temporal dynamic sparsity & {${\cal O}\left(TKL+TK+(s_p+j)^3\right)$}\\ 
\hline
\end{tabular}}
\end{center}
\footnotesize{Notes: $q$ is the modulation order, $w$ is the maximum number of symbols spreading over the same subcarrier, $s_p$ is the quality information, and $j$ is the update index of sparsity level.}
\label{Tab1}
\vspace*{-0mm}
\end{table*}

\begin{figure}[!t]
\begin{center}
\includegraphics[width=0.85\columnwidth, keepaspectratio]{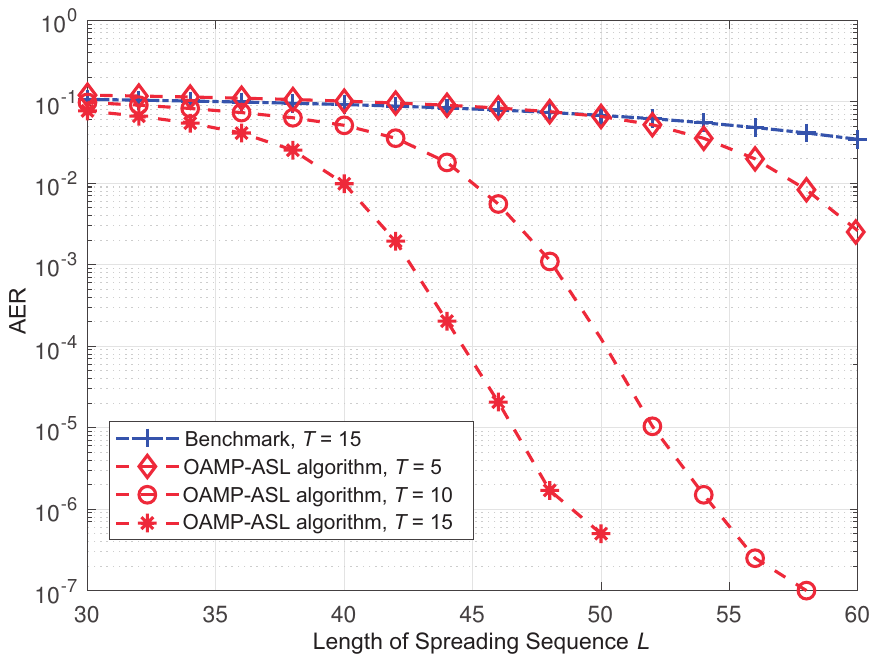}
\end{center}
\caption{\small{AER performance comparison of the advanced OAMP-ASL-based JADD scheme \cite{Mei2022JADD} and the benchmark \cite{Determe2017SOMP}, where $K = 500$ and $K_a = 50$, and $K_a$ is unknown at the BS.}}     
\label{Fig6}
\end{figure}

Fig.~\ref{Fig6} provides an example to verify the superiority of leveraging the temporal common sparsity, where the activity error rate (AER) of the OAMP-ASL algorithm \cite{Mei2022JADD}, which incorporates the temporal common sparsity, is compared with that of the traditional greedy CS recovery algorithm \cite{Determe2017SOMP} without considering the temporal common sparsity. As can be observed, the OAMP-ASL algorithm considerably outperforms the baseline scheme without incorporating the temporal common sparsity. Also as expected, the performance of the OAMP-ASL algorithm improves as the number of symbols within a slot $T$ increases, since a larger $T$ indicates an enhanced temporal common sparsity.

\begin{figure}[!t]
\begin{center}
\includegraphics[width=0.85\columnwidth, keepaspectratio]{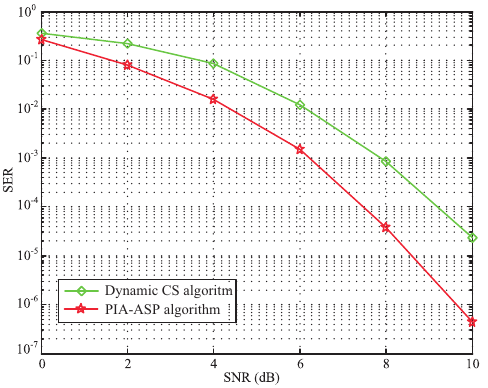}
\end{center}
\caption{\small{SER performance comparison of the dynamic CS-based JADD scheme and PIA-ASP-based JADD scheme, where $K = 200$, $L = 100$, and $K_a$ varies from 14 to 20 in 7 different time slots \cite{Du2017JADD} \textcircled{c}IEEE.}}     
\label{Fig7}
\end{figure}

On the other hand, focusing on the temporal dynamic sparsity, the authors in \cite{Wang2016DCS} proposed a dynamic CS-based JADD approach, where the active device set estimated in the current slot is adopted to initialize the estimate of the active device set in the next slot. However, the solution of \cite{Wang2016DCS} assumes the availability of the sparsity level, i.e., the number of active devices, which is unrealistic in practical scenarios. Moreover, the prior information is exploited blindly where the reliability of the estimate from the previous slot is not evaluated. To overcome this limitation, a prior-information aided adaptive subspace pursuit (PIA-ASP) algorithm \cite{Du2017JADD} is developed, which reaps a better symbol error rate (SER) performance, as illustrated in Fig.~\ref{Fig7}. 

For the benefit of the reader, a brief summary of the aforementioned JADD schemes is provided in Table~\ref{Tab1}.

\subsubsection{JADD for GFMA in Massive MIMO Systems}\label{S4.1.2}

Massive MIMO has been identified as a pivotal technique for current 5G NR and future beyond 5G (B5G)/{6G} cellular systems, providing game-changing improvements in the spectral and energy efficiencies \cite{Marzetta2010mMIMO, Ng2012mMIMO, Ngo2013mMIMO}. Moreover, the transmission reliability of massive IoT connectivity can be considerably improved by leveraging the extra spatial DoF \cite{Bana2019mMIMO}. {To reap these benefits, one effective way is to intuitively extend the problem formulation in the previous subsection to massive MIMO systems.} Specifically, for massive MIMO systems with $N_r$ BS antennas, the channel vector between the BS and the $k$th device, denoted by ${\bf h}_k \in \mathbb{C}^{N_r \times 1}$, is unique for the device, which can be regarded as a device-specific signature spreading in the spatial domain. Naturally, if the channels associated with all the $K$ devices are available at the BS, the JADD for GFMA can be formulated as a MMV CS problem of (\ref{Eq1}). Here, ${\bf \Phi} = \left[{\bf h}_1, {\bf h}_2, \cdots, {\bf h}_K\right] \in \mathbb{C}^{N_r \times K}$ is the known massive access channel matrix and ${\bf X} = \left[\alpha_1{\bf x}_1, \alpha_2{\bf x}_2, \cdots, \alpha_K{\bf x}_K\right]^{\rm T} \in \mathbb{C}^{K \times T}$ is the sparse signal matrix as explained in Subsection~\ref{S4.1.1}. In this context, the CS recovery algorithms introduced in Subsection~\ref{S4.1.1} can be directly applied to leverage the temporal common sparsity or temporal dynamic sparsity of ${\bf X}$.

\begin{figure}[!t]
\begin{center}
\includegraphics[width=1\columnwidth, keepaspectratio]{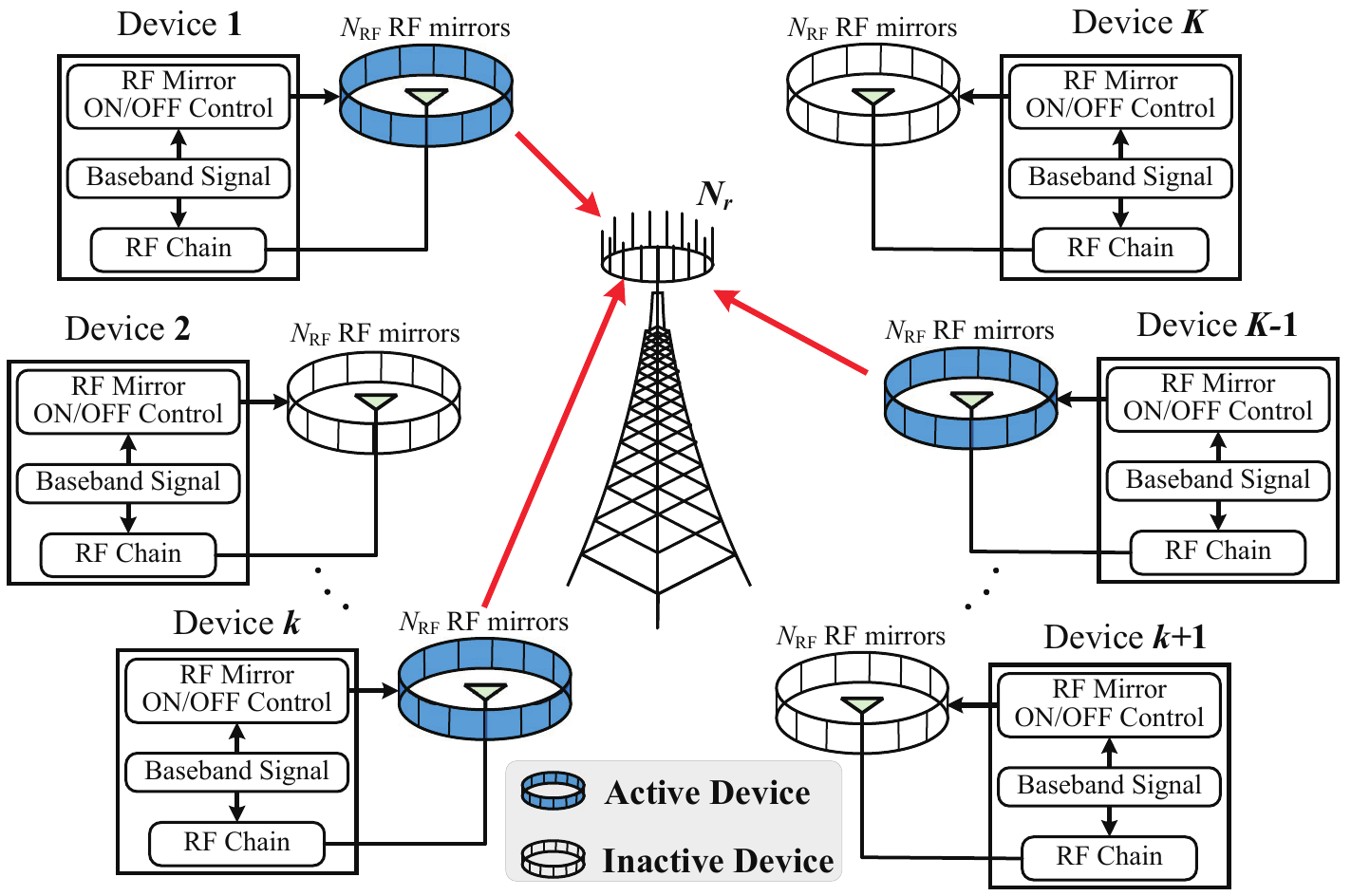}
\end{center}
\caption{\small{The system model and transmitter structure of MBM-based GFMA scheme in massive MIMO systems \cite{Qiao2022MBM} \textcircled{c}IEEE.}}
\label{Fig8}
\end{figure}

On the other hand, by equipping multiple antennas at the devices, the spatial modulation (SM) can be incorporated to boost the spectral efficiency for GFMA, without increasing the hardware complexity and energy consumption of the devices~\cite{Gao2016SM, Meng2016SM}. In the SM scheme, each active device activates only one transmit antenna, based on which the additional ${\rm log}_2\left(N_t\right)$-bit information can be conveyed through the active antenna index \cite{Ma2019SM,IndexMod}, where $N_t$ is the number of transmit antennas. SM is a so-called index modulation scheme that exploits the transmit antenna index to convey additional information bits \cite{IndexMod}. In this context, only one radio-frequency (RF) chain is required at the devices, but the number of transmit antennas scales exponentially with the number of additionally conveyed information bits. Following a similar index modulation idea, the more efficient media-based modulation (MBM) \cite{Khandani2013MBM, Shamasundar2018MBM, Zhang2020MBM} has been widely investigated to overcome the aforementioned limitation of SM. Specifically, each device is equipped with one RF chain, one transmit antenna, and $N_{\rm RF}$ low-cost RF mirrors, where each RF mirror has a controllable binary ON/OFF status, as illustrated in Fig.~\ref{Fig8}. Therefore, each device has $N_t = 2^{N_{\rm RF}}$ different mirror activation patterns, i.e., $N_t$ different channel realizations, which can be exploited to encode $N_{\rm RF}$ extra information bits.

\begin{figure}[!t]
\begin{center}
\includegraphics[width=1\columnwidth, keepaspectratio]{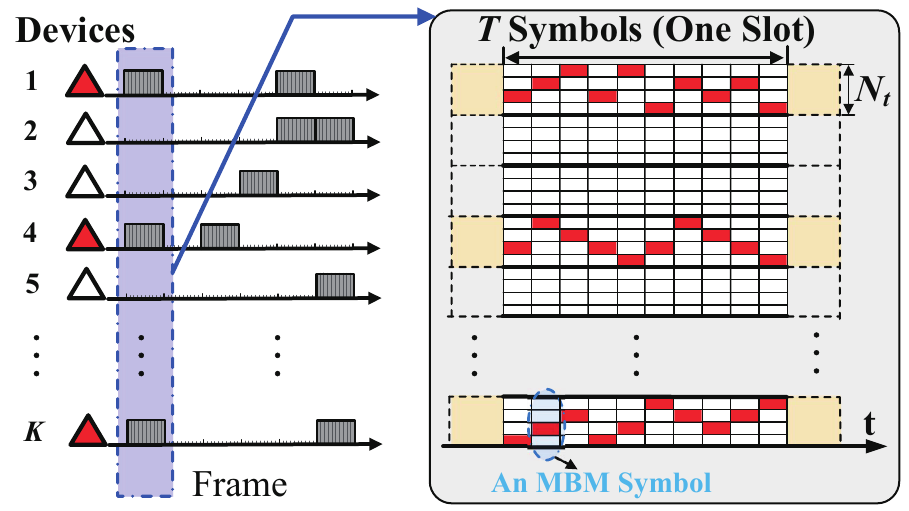}
\end{center}
\caption{\small{Doubly structured sparsity of the access signal matrix in MBM-based GFMA scheme \cite{Qiao2022MBM} \textcircled{c}IEEE.}
\label{Fig9}}
\end{figure}

For both the SM-based and MBM-based GFMA schemes, the related JADD at the BS can also be formulated as a MMV CS problem as expressed in (\ref{Eq1}). Specifically, the sensing matrix is ${\bf \Phi} = \left[{\bf H}_1, {\bf H}_2, \cdots, {\bf H}_K\right] \in \mathbb{C}^{N_r \times KN_t}$ with ${\bf H}_k \in \mathbb{C}^{N_r \times N_t}$ denoting the MIMO channel matrix between the BS and the $k$th device for all the channel realizations. Again assume that the full CSI is available at the BS and the device activity remains constant within a slot having $T$ successive symbols. The $t$th column of the sparse signal matrix ${\bf X}$ is expressed as $\left[{\bf X}\right]_{:,t} = \left[\left({\bf x}_{1,t}\right)^{\rm T}, \left({\bf x}_{1,t}\right)^{\rm T}, \cdots, \left({\bf x}_{K,t}\right)^{\rm T} \right]^{\rm T} \in \mathbb{C}^{KN_t \times 1}$, where ${\bf x}_{k,t}  = \alpha_k{s_{k,t}}{\bf d}_{k,t} \in \mathbb{C}^{N_t \times 1}$ is the uplink access signal of the $k$th device transmitted in the $t$th symbol duration. Here, $\alpha_k \in \left\{0, 1\right\}$, $\forall k \in \left\{1, 2, \cdots, K\right\}$, denotes the activity indicator, $s_{k,t} \in \mathbb{C}$ is the conventional modulated symbol, and ${\bf d}_{k,t} \in \mathbb{C}^{N_t \times 1}$ is the MBM vector which has unity on the index corresponding to the activated RF mirror and zeros elsewhere. Note that the total information bits are encoded in both the modulated symbol $s_{k,t}$ and the non-zero index of ${\bf d}_{k,t}$. Due to the sporadic uplink traffic of devices and the characteristics of MBM, the signal matrix ${\bf X}$ exhibits \emph{doubly structured sparsity}, as illustrated in Fig. \ref{Fig9}. Specifically, in each column of ${\bf X}$, only the access signals of $K_a$ active devices are non-zero, and all columns share the same device level sparsity. Moreover, the access signal of a specific active device ${\bf x}_{k,t}$ is also sparse, where only one entry of the MBM vector ${\bf d}_{k,t}$ is unity and the others are zero. 

To exploit the doubly structured sparsity for improving JADD performance, a variety of CS recovery algorithms have been developed. Inspired by the idea of subspace matching pursuit, the authors in \cite{Ma2019SM} proposed a greedy two-level structured sparsity (TLSS)-based detector for MBM-based GFMA. Subsequently, a two-stage detection scheme was further developed, where a structured OMP (StrOMP) algorithm was proposed for activity detection and an SIC-based structured SP (SIC-SSP) algorithm was designed for the demodulation of the detected active devices \cite{Qiao2020MBM}, which will be denoted as StrOMP+SIC-SSP. The aforementioned works focus on only a single time slot in which the device activity remains constant over multiple successive symbol durations. Furthermore, the authors in \cite{Ma2020MBM} proposed a prior-information aided adaptive media modulation subspace matching pursuit (PIA-MSMP) algorithm to accommodate the dynamic device activity across different time slots within a frame. Different from the greedy CS recovery algorithms proposed in \cite{Ma2019SM, Qiao2020MBM, Ma2020MBM}, a doubly structured AMP (DS-AMP) algorithm was developed under the Bayesian framework, which further takes the {\em a priori} information of the finite constellations of ${\bf X}$ into account \cite{Qiao2022MBM}. Moreover, the theoretical state evolution (SE) of the DS-AMP algorithm was derived to analyze its performance in \cite{Qiao2022MBM}. Compared to single-antenna systems, the channel estimation in MBM-based massive MIMO systems is much more challenging. Therefore, the GFMA schemes introduced in this subsection are mainly tailored for the IoT applications where the devices are fixed or have very low mobility, thus the CSI can be estimated accurately and it does not have to be updated frequently. In particular, the channel estimation issue for media modulation-based GFMA was also investigated in \cite{Qiao2022MBM} and \cite{Ma2020MBM}. A brief summary of the aforementioned representative JADD algorithms for MBM-based GFMA is provided in Table~\ref{Tab2}. 


\begin{table*}[t!] 
\caption{\small{Summary of JADD algorithms for MBM-based GFMA scheme in massive MIMO systems}}
\begin{center}
\resizebox{\linewidth}{!}{ 
\begin{tabular}{|l|c|c|l|l|l|}
\hline
\multicolumn{1}{|c|}{JADD Scheme}	&	CS Model	&	CS Algorithm	&	\multicolumn{1}{c|}{Sparsity Structure}		&	\multicolumn{1}{c|}{Advances} & {Complexity}\\ 
\hline
TLSS-based algorithm \cite{Ma2019SM}		&	MMV			&	Greedy			&	Doubly structured sparsity					&	Exploit the doubly structured sparsity & {${\cal O}\left(K_a^3+N_r^2+N_t^3\right)$}\\ 
\hline
StrOMP SIC-SSP algorithm \cite{Qiao2020MBM}	&	MMV			&	Greedy			&	Doubly structured sparsity					&	\begin{tabular}[c]{@{}l@{}} Propose a two-stage detection scheme and \\ 
employ SIC for enhanced performance \end{tabular} & {${\cal O}\left(K_a^3\right)$}\\ 
\hline
PIA-MSMP algorithm \cite{Ma2020MBM}			&	MMV			&	Greedy			&	\begin{tabular}[c]{@{}l@{}} Doubly structured sparsity and \\ dynamic device activity \end{tabular}		&	Further consider the dynamic device activity & {${\cal O}\left(TK_aN_rN_t+N_r^2+N_t^3\right)$}\\ 
\hline
DS-AMP algorithm \cite{Qiao2022MBM}			&	MMV			&	Bayesian		&		Doubly structured sparsity				&	\begin{tabular}[c]{@{}l@{}} Further exploit the statistical information \\ of the access signal matrix\end{tabular} & {${\cal O}\left(TKN_rN_t\right)$}\\ 
\hline
\end{tabular}}
\end{center}
\label{Tab2}
\end{table*}

\begin{figure*}[!t] 
\begin{center}
\subfloat[]{\includegraphics[width=2.9 in]{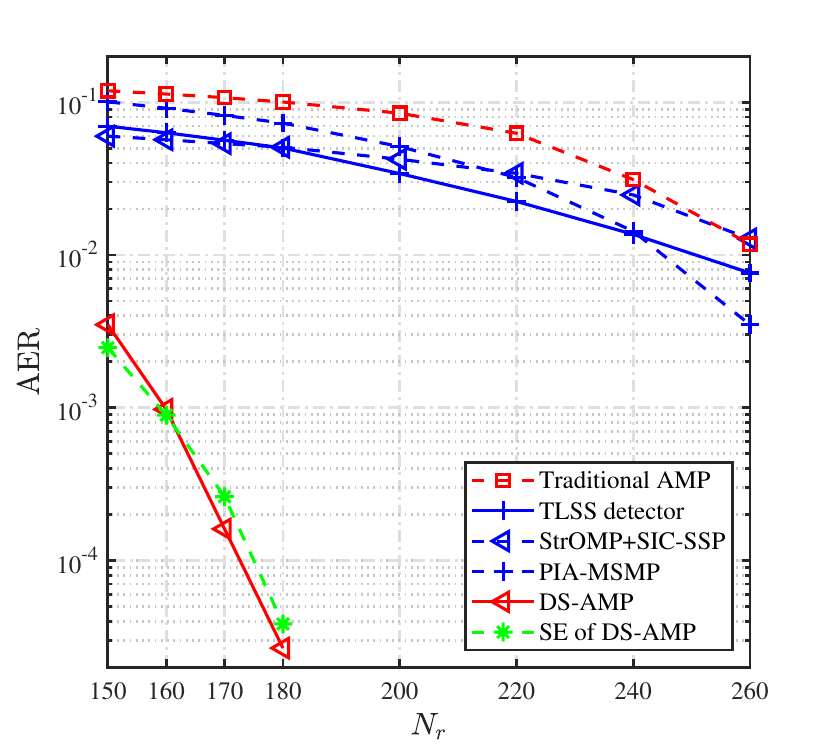}}
\hspace{5mm}
\subfloat[]{\includegraphics[width=2.9 in]{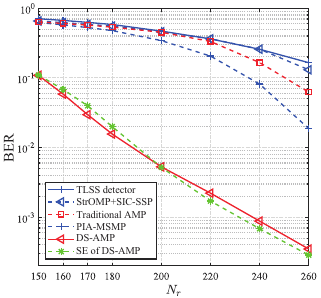}}
\end{center}
\caption{\small{{Performance comparison of different JADD algorithms for MBM-based GFMA schemes, where $K = 500$, $K_a = 50$, $N_{\rm RF} = 2$, and $T = 12$ are considered: (a)~AER performance, and (b)~BER performance \cite{Qiao2022MBM} \textcircled{c}IEEE.}}}
\label{Fig10}
\end{figure*}

Fig.~\ref{Fig10} provides an example to compare the JADD performance of the discussed algorithms as well as the traditional AMP algorithm \cite{Dohono2010AMP}, in terms of both AER and bit error rate (BER), where `SE of DS-AMP' represents the theoretical SE of the DS-AMP. Obviously, by fully exploiting the doubly structured sparsity and the {\em a priori} statistical information of ${\bf X}$, the DS-AMP algorithm significantly outperforms its counterparts that do not leverage the structured sparsity or do not leverage the {\em a priori} statistical information. Moreover, both AER and BER performance becomes better as the number of BS antennas increases, which verifies the superiority of massive MIMO in MBM-based GFMA. Besides, the derived SE can accurately predict the performance of the DS-AMP algorithm, providing insightful guidance for optimizing practical system designs.

\subsection{Joint Activity Detection and Channel Estimation}\label{S4.2}
 
It should be noted that the GFMA cannot always be formulated as a JADD problem. This is because the JADD is based on the condition that the full CSI is available at the BS. In many IoT applications, such as smart traffic and wearable IoT, the channels between the devices and the BS may change frequently. In this context, it is unrealistic to assume that the full CSI is available at the BS, especially for massive MIMO systems with a massive number of devices. Therefore, the frame structure with the format of `pilot + data' has recently been proposed, where each device is assigned with a unique non-orthogonal pilot sequence for joint activity detection and channel estimation (JADCE) at the BS. With the estimated active device set and the corresponding channels, the conventional coherent data detection is then executed based on the received data signal \cite{Liu2018mMTC}. Specifically, the JADCE problem can be formulated as
\begin{equation}\label{Eq2}
 {\bf Y} = {\bf P}{\bf H} + {\bf N},
\end{equation}
where ${\bf Y}\in \mathbb{C}^{P \times N_r}$ is the received pilot signal, ${\bf P} = \left[{\bf p}_1, {\bf p}_2, \cdots, {\bf p}_K\right] \in \mathbb{C}^{P \times K}$ is the pilot matrix, and ${\bf p}_k \in \mathbb{C}^{P \times 1}$ is the non-orthogonal pilot sequence of the $k$th device with the pilot length $P$, while ${\bf H} = \left[\alpha_1{\bf h}_1, \alpha_2{\bf h}_2, \cdots, \alpha_K{\bf h}_K\right]^{\rm T} \in \mathbb{C}^{K \times N_r}$ is the massive access channel matrix, $\alpha_k$ is again the activity indicator of the the $k$th device, and ${\bf N}$ is the AWGN. Considering the sporadic uplink traffic of devices, the JADCE problem (\ref{Eq2}) becomes an SMV CS problem in single-antenna systems, i.e., $N_r = 1$, and an MMV CS problem in MIMO systems, i.e., $N_r > 1$.

In \cite{Schepker2013JADCE} and \cite{Ahn2019JADCE}, the authors proposed two CS-based JADCE schemes, respectively, for GFMA in single-antenna systems, where the OMP-based and EP-based CS recovery algorithms are developed, respectively. Furthermore, the authors in \cite{Liu2018JADCE2} revealed that the error probability of device activity detection can be made arbitrary small by increasing the number of BS antennas. {Based on this attractive finding, a large number of JADCE schemes have been proposed in single-station massive MIMO systems \cite{Liu2018JADCE2, Knoop2016JADCE, Park2017JADCE, Liu2018JADCE3, Shao2018JADCE, Chen2018JADCE, Shao2020JADCE, Jiang2021JADCE, Ke2020JADCE, Jiang2022JADCE}, and then naturally extended to more complicated cooperative massive MIMO systems \cite{Ngo2017CF, Xu2015CRAN, Utkovski2016CRAN, Chen2019MC, Ke2020CF}. Following this line,} we will first discuss the JADCE problem in single-station massive MIMO systems and then extend it to cooperative massive MIMO systems.

\begin{figure}[!t]
\includegraphics[width=1\columnwidth, keepaspectratio]{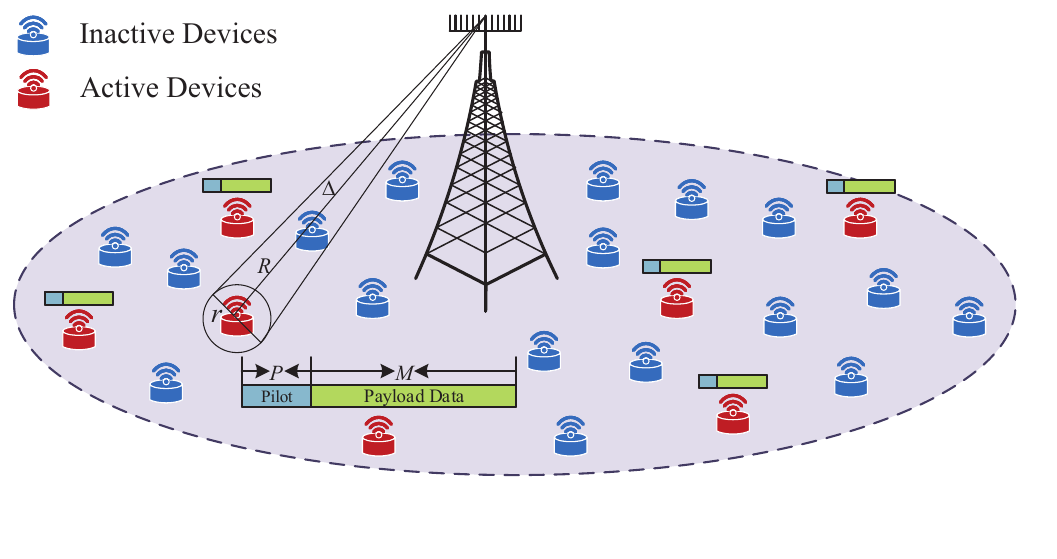}
\vspace*{-8mm}
\caption{\small{The system model of JADCE for GFMA in single-station massive MIMO systems.}}
\label{Fig11}
\end{figure}

\subsubsection{JADCE in Single-Station Massive MIMO Systems}\label{S4.2.1}

\begin{table*}[t!] 
\caption{\small{Summary of JADCE schemes for GFMA in single-station massive MIMO systems}}
\begin{center}
\resizebox{\linewidth}{!}{  
\begin{tabular}{|l|l|l|l|l|}
\hline
\multicolumn{1}{|c|}{Reference}							&		\multicolumn{1}{c|}{Channel Model}		&		\multicolumn{1}{c|}{Advances}  & {Complexity} \\ 
\hline
L. Liu, \emph{et al.} \cite{Liu2018JADCE2}				& 		Spatial domain          				& 		\begin{tabular}[c]{@{}l@{}}	Leverage the spatial-domain common sparsity across different BS antennas \end{tabular} & {${\cal O}\left(PKN_r\right)$}\\ 
\hline
L. Liu, \emph{et al.} \cite{Liu2018JADCE3}        		& 		Spatial domain             				& 		\begin{tabular}[c]{@{}l@{}}	Analyze the uplink achievable rate and optimize the pilot length \end{tabular} & {N/A} \\ 
\hline
X. Shao, \emph{et al.} \cite{Shao2018JADCE}         	& 		Spatial domain                   		& 		\begin{tabular}[c]{@{}l@{}}	Propose a three-phase unified transmission design for GFMA \end{tabular} & {${\cal O}\left(PKN_r\right)$}\\
\hline
Z. Chen, \emph{et al.} \cite{Chen2018JADCE}         	& 		Spatial domain          				& 		\begin{tabular}[c]{@{}l@{}}	Consider the unknown large-scale fading parameter \end{tabular} & {${\cal O}\left(PKN_r\right)$}\\
\hline
X. Shao, \emph{et al.} \cite{Shao2020JADCE}         	& 		Spatial domain         					& 		\begin{tabular}[c]{@{}l@{}}	Propose a dimension-reduced algorithm to reduce the computational complexity \end{tabular} & {${\cal O}\left(PN_rr_e+r_e^3\right)$} \\ 
\hline
J. Jiang, \emph{et al.} \cite{Jiang2021JADCE}        	& 		Spatial domain           				& 		\begin{tabular}[c]{@{}l@{}}	Consider the time-varying device activity and CSI \end{tabular} &{${\cal O}\left(PKN_r\right)$} \\
\hline
M. Ke, \emph{et al.} \cite{Ke2020JADCE}           		& 		\begin{tabular}[c]{@{}l@{}} Spatial domain $\&$ \\ Angular domain \end{tabular}		&		\begin{tabular}[c]{@{}l@{}} Leverage both the spatial-domain common sparsity and the angular-domain \\ clustered sparsity \end{tabular} & {${\cal O}\left(PKN_rN_c\right)$} \\ 
\hline
J. Jiang, \emph{et al.} \cite{Jiang2022JADCE}        	& 		Angular domain      					& 		\begin{tabular}[c]{@{}l@{}}	Leverage the fact that some devices may have  common local scattering cluster \end{tabular} & {${\cal O}\left(PK_gu_g\right)$} \\ 
\hline
\end{tabular}}
\end{center}
\label{Tab3}
Notes: $r_e$ is the rank for dimension reduction, $N_c$ is the number of subcarriers, $K_g$ is he number of devices in group $g$, and $u_g$ is the sparsity of sparsity level of angular-domain channel matrix.
\end{table*}

A BS equipped with an $N_r$-elements uniform linear array (ULA) serves $K$ single-antenna devices distributed in its coverage, where only $K_a$ devices are activated in each frame duration, as illustrated in Fig.~\ref{Fig11}. A two-phase transmission scheme is adopted, with each frame consisting of a pilot phase and the subsequent payload data phase. Each device is assigned with a unique non-orthogonal pilot sequence that will be transmitted in the pilot phase. When accessing the network, the active devices directly transmit their uplink access signals with the format of `pilot + data' on the same time-frequency resources. In the pilot phase, given the received pilot signal ${\bf Y}$ and the pre-allocated pilot matrix ${\bf P}$, the JADCE problem is equivalent to estimating the sparse channel matrix ${\bf H}$ based on the MMV CS model of (\ref{Eq2}). Following this formulation, an OMP-based JADCE scheme was developed for GFMA in single-station massive MIMO systems \cite{Knoop2016JADCE}, and the sparsity of the delay-domain CIR was further leveraged to improve the channel estimation accuracy \cite{Park2017JADCE}.

On the other hand, equipping a large number of antennas at the BS results in additional sparsity properties of the massive access channel matrix, which can be leveraged to further enhance JADCE performance.  Specifically, the sporadic traffic of devices leads to the sparsity of the channel vector associated with each  receive antenna, i.e., every column of ${\bf H}$ is sparse. Moreover, all the BS antennas observe a common sparsity pattern, which leads to the spatial-domain common sparsity of the channel matrix and facilitates the activity detection through non-zero row detection \cite{Liu2018mMTC}. Based on the spatial-domain signal model (\ref{Eq2}), several efficient JADCE schemes were proposed \cite{Liu2018JADCE2, Liu2018JADCE3, Shao2018JADCE, Chen2018JADCE, Shao2020JADCE, Jiang2021JADCE}. Specifically, in \cite{Liu2018JADCE2}, a vector AMP algorithm was developed to exploit the common sparsity across different BS antennas, and the related probabilities of false alarm and miss detection were analyzed exploiting the SE. Adopting this vector AMP algorithm, each active device's uplink achievable rate was further characterized, based on which the length of the non-orthogonal pilot sequence was optimized in \cite{Liu2018JADCE3}. Afterward, the authors in \cite{Shao2018JADCE} designed a three-phase transmission protocol for GFMA, which also employed the vector AMP algorithm and further considered the downlink transmission phase. The works \cite{Liu2018JADCE2, Liu2018JADCE3, Shao2018JADCE} assume that the large-scale component of the channel fading coefficients are known to the BS. Considering a more practical scenario, an updated vector AMP algorithm was derived in \cite{Chen2018JADCE}, which takes the unknown large-scale fading parameters into account. Due to the large numbers of devices and BS antennas, the JADCE generally imposes a high computational complexity. To mitigate this problem, a dimension reduction-based JADCE scheme was proposed in \cite{Shao2020JADCE}, which projects the original channel matrix onto a low-dimensional space by jointly exploiting its sparse and low-rank structures. In addition, considering the time-varying device activity and CSI, the work \cite{Jiang2021JADCE} claimed that the inherent temporal correlation between adjacent time slots can be exploited to enhance the JADCE performance. 

\begin{figure*}[!h]
\begin{center}
\subfloat[]{\includegraphics[width=2.9 in]{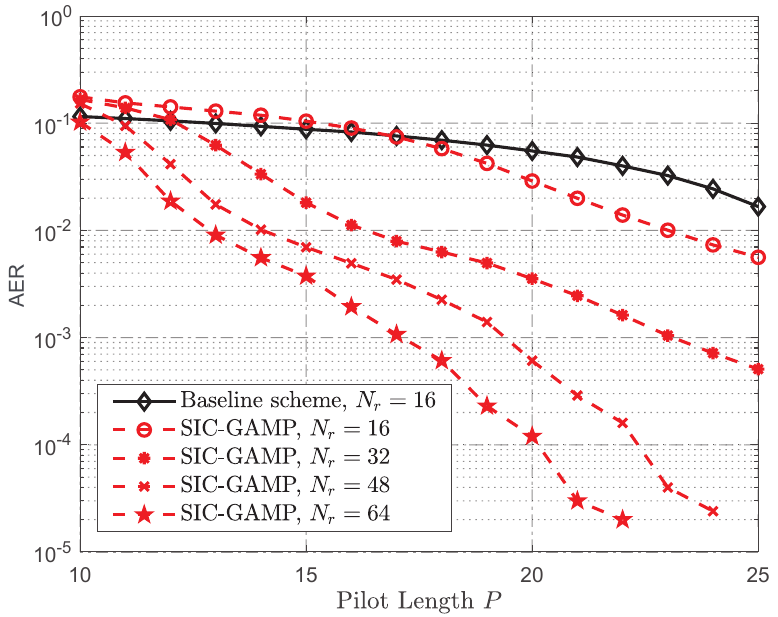}}
\hspace{5mm}
\subfloat[]{\includegraphics[width=2.9 in]{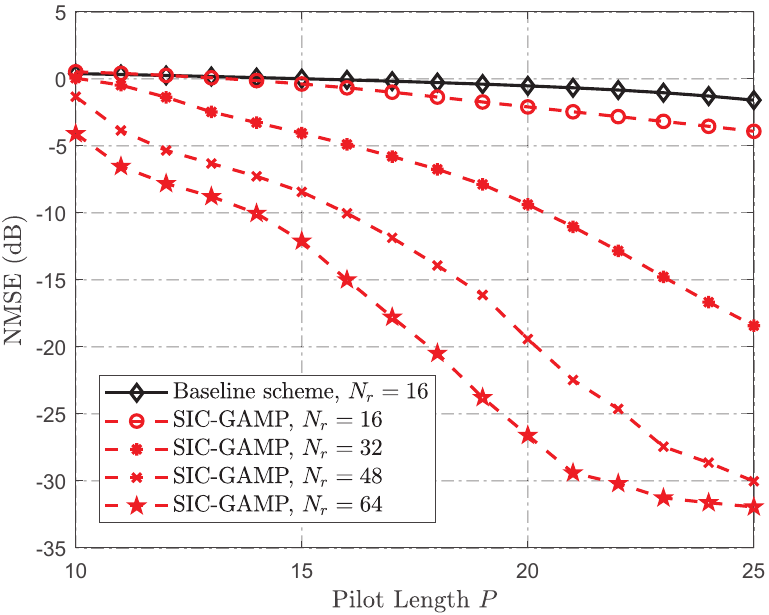}}
\end{center}
\caption{\small{Comparison JADCE performance of the SIC-GAMP \cite{Ke2020JADCE} and the baseline scheme \cite{Liu2018JADCE2} for GFMA in single-station massive MIMO systems: (a)~AER performance, and (b)~NMSE performance.}}
\label{Fig12}
\end{figure*}

In \cite{Liu2018JADCE2, Liu2018JADCE3, Shao2018JADCE, Chen2018JADCE, Shao2020JADCE, Jiang2021JADCE}, the reduction of the pilot overhead is limited to the number of active devices, i.e., $P \ge K_a$ is required, which becomes a severe obstacle for the implementation of these spatial-domain JADCE schemes. Further considering the typical one-ring channel model, the massive MIMO channel vectors exhibit clustered sparsity in the angular domain \cite{Gao2018CS}. Hence, the JADCE problem (\ref{Eq2}) can be re-formulated as
\vspace*{-1mm}
\begin{equation}\label{Eq3}
{\bf R} = {\bf P}{\widetilde {\bf H}} + {\widetilde {\bf N}},
\vspace*{-1mm}	
\end{equation}
where ${\bf R} = {\bf Y} {\bf A}_R$ is the angular-domain received signal, ${\widetilde {\bf H}}  = {\bf H}{\bf A}_R$ is the angular-domain channel matrix, ${\widetilde {\bf N}} = {\bf N}{\bf A}_R$, and ${\bf A}_R$ is the spatial-to-angular domain transformation matrix determined by the geometrical structure of the BS array. In contrast to the spatial-domain channel model (\ref{Eq2}), the angular-domain channel model (\ref{Eq3}) is more favorable to improve the accuracy of the CSI estimates of the identified active devices. Motivated by this angular-domain channel model, an SIC-based generalized AMP (GAMP) algorithm was proposed to jointly exploit the spatial-domain and angular-domain structured sparsities, where the pilot overhead can be far smaller than the number of active devices~\cite{Ke2020JADCE}. Considering that some devices may experience common local scattering clusters, a grouping-based JADCE scheme was proposed in \cite{Jiang2022JADCE}. 

\begin{figure*}[!t]
\begin{center}
\subfloat[]{\includegraphics[width=2.9 in]{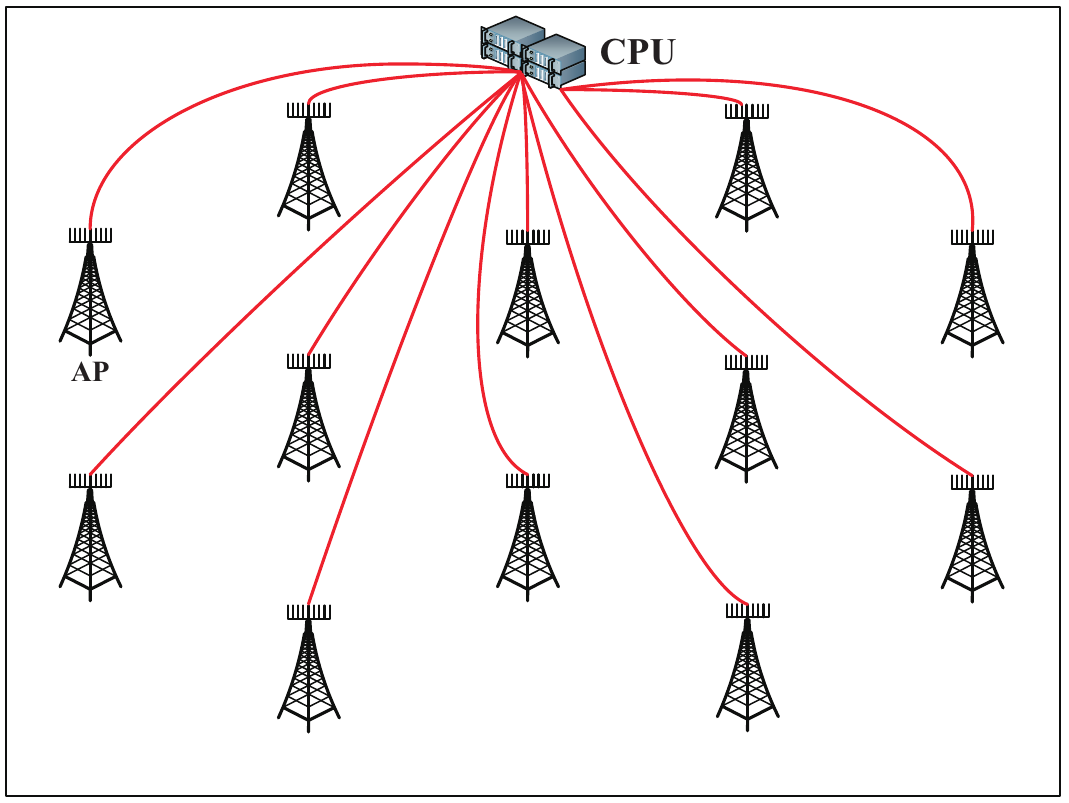}}\hspace{5mm}
\subfloat[]{\includegraphics[width=2.9 in]{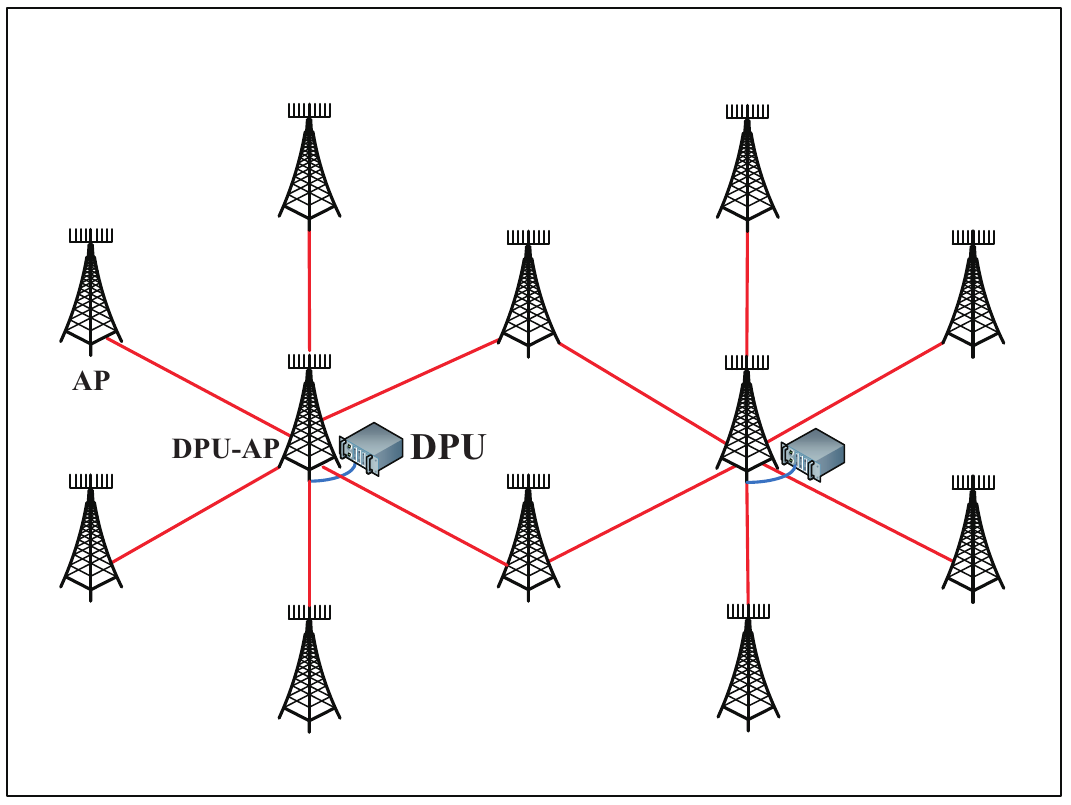}}
\end{center}
\caption{\small{Two processing paradigms in cell-free massive MIMO systems: (a)~cloud computing, and (b)~edge computing \cite{Ke2020CF} \textcircled{c}IEEE.}}
\label{Fig13}
\end{figure*}

A brief summary of the aforementioned JADCE schemes is provided in Table~\ref{Tab3}.  Fig.~\ref{Fig12} verifies the effectiveness of the SIC-GAMP algorithm in the case of a single BS massive MIMO. Here, $K = 500$, $K_a = 50$, and SNR $= 20$~dB are considered, and the numbers of BS antennas is set to $N_r = 16$, 32, 48, or $64$.  The  state-of-the-art JADCE scheme \cite{Liu2018JADCE2} that only considers the spatial-domain model (\ref{Eq2}) is adopted as the baseline scheme. It is observed that by exploiting the angular-domain clustered sparsity of massive MIMO channels, the SIC-GAMP scheme attains a significant performance improvement over the baseline scheme. Furthermore, the achievable performance of the SIC-GAMP scheme improves with the increase of  the number of BS antennas. This is because increasing the number of BS antennas can simultaneously enhance the spatial-domain common sparsity of ${\bf H}$ and the angular-domain clustered sparsity of ${\widetilde {\bf H}}$, which improves the CS recovery performance. In particular, the SIC-GAMP scheme can reliably support GFMA even at an overloading ratio of $250~\%$ (i.e., $P = 20$ and $K_a = 50$). With $N_r=64$, the scheme achieves an AER of $10^{-4}$ and a normalized mean square error (NMSE) of $-27$ dB.

\subsubsection{JADCE in Cooperative Massive MIMO Systems}\label{S4.2.2}

For typical IoT applications, the power-limited devices are generally distributed in a vast area, and thus multiple BSs should be densely deployed to offer an adequate coverage and save the transmit power of the devices. Adopting the traditional small-cell massive MIMO networks, the reduced BS spacing however would inevitably introduce severe uplink inter-cell interferences, which is a limiting factor for reliable GFMA \cite{Ngo2017CF}. To overcome this limitation, various cooperative massive MIMO networks have been intensively investigated for GFMA. The authors in \cite{Xu2015CRAN} extended the JADCE problem of GFMA to the cloud radio access network (C-RAN), where the received signals from all the BSs are jointly processed at a central unit. The authors in \cite{Utkovski2016CRAN} further considered the limited capacity of the backhaul links between the BSs and the central unit. Moreover, the authors in \cite{Chen2019MC} studied the JADCE of GFMA in multi-cell systems, and compared the conventional non-cooperative massive MIMO network and the cooperative massive MIMO network in terms of their effectiveness in overcoming inter-cell interferences.

\begin{table*}[t!]
\caption{\small{Summary of JADCE schemes for GFMA in cooperative massive MIMO systems}} 
\begin{center}
\resizebox{\linewidth}{!}{  
\begin{tabular}{|l|l|l|l|}
\hline
\multicolumn{1}{|c|}{Reference}							&		\multicolumn{1}{c|}{Network Architecture}		&		\multicolumn{1}{c|}{Advances} &{ Complexity}\\ 
\hline
X. Xu, \emph{et al.} \cite{Xu2015CRAN}            		& 		C-RAN                                     		& 		Extend the JADCE problem to the C-RAN & {${\cal O}\left(P^2+N_r^2\right)$} \\
\hline
Z. Utkovski, \emph{et al.} \cite{Utkovski2016CRAN}     	& 		C-RAN                         					& 		Consider the limited capacity of the backhaul links & {${\cal O}\left(PKN_rN_p\right)$} \\
\hline
Z. Chen, \emph{et al.} \cite{Chen2019MC}          		& 		\begin{tabular}[c]{@{}l@{}} Multi-cell massive MIMO \\ and cooperative massive MIMO \end{tabular}		&		\begin{tabular}[c]{@{}l@{}} Compare the JADCE performance of multi-cell massive \\ MIMO and cooperative massive MIMO \end{tabular} & {${\cal O}\left(PKN_rN_p\right)$}  \\
\hline
M. Ke, \emph{et al.} \cite{Ke2020CF}             		& 		Cell-free massive MIMO            				& 		\begin{tabular}[c]{@{}l@{}} Propose cloud computing and edge computing paradigms \\ for signal processing \end{tabular} & {${\cal O}\left(PKN_rN_cN_p\right)$} \\
\hline
\end{tabular}}
\end{center}
Notes: $N_c$ is the number of subcarriers and $N_p$ is the number of APs.
\label{Tab4}
\end{table*}

Among various cooperative massive MIMO networks, cell-free massive MIMO networks are the most popular ones and have attracted ever increasing attention from both the academic and industrial communities \cite{Ngo2017CF}. In fact, cell-free massive MIMO is an incarnation of the general idea of distributed MIMO, network MIMO, C-RAN, and distributed antenna systems, where a large number of access points (APs) cooperate with each other in the network to serve a large area. The APs equipped with a large number of antennas are connected to the related processing units via fronthaul links for joint signal processing, as illustrated in Fig.~\ref{Fig13}. In this context, the inter-cell interference can be effectively avoided, as cells and cell boundaries do not exist. Moreover, by performing coherent signal processing across geographically distributed APs' antennas, cell-free massive MIMO can provide a uniformly good service for all devices. By contrast, for centralized massive MIMO systems with the receive antennas locating at BSs, the devices in the cell center generally reaps a better service quality than the devices in the cell edge due to the heterogeneous path loss effect \cite{Xu2015CRAN}.  Besides, equipping massive antennas at the APs further combines the distributed MIMO and massive MIMO concepts, which is expected to reap all the benefits from these two systems.

Based on the notion of cell-free massive MIMO networks, two different signal processing paradigms, namely, cloud computing of Fig.~\ref{Fig13}\,(a) and edge computing of Fig.~\ref{Fig13}\,(b), have been proposed for supporting centralized and distributed AP cooperation, respectively \cite{Ke2020CF}. For cloud computing, the signals received at all the APs are centrally processed in a central processing unit (CPU). Since the APs are only designed to work as relays with simple signal processing capabilities only, the required cost for large scale deployment of APs is significantly reduced. However, cloud computing requires the information to pass through several sub-networks including the radio access network, backhaul network, and core network, where traffic control, routing, and other network-management operations can contribute to excessive delays. As for edge computing, the central processing is offloaded to some of the APs equipped with distributed processing units (DPUs) such that the corresponding computations can be performed in a distributed manner. Compared to cloud computing, edge computing can alleviate the burden on the fronthaul links and the CPU, facilitate a faster access response as well as support more efficient AP cooperation, at the expense of the increased cost in network deployment \cite{Ke2020CF}.

\begin{figure}[!t]
\begin{center}
\includegraphics[width=0.85\columnwidth, keepaspectratio]{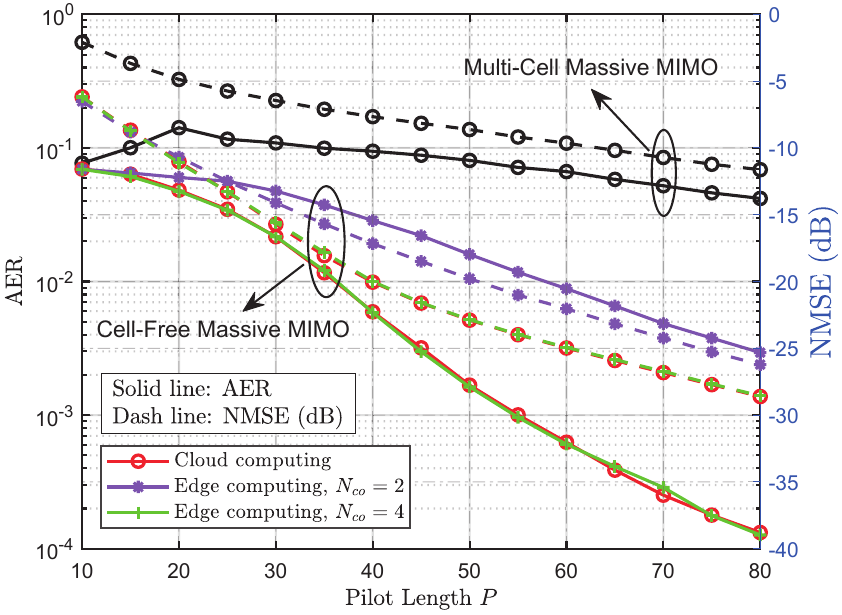}
\end{center}
\caption{\small{JADCE performance of GFMA in multi-cell non-cooperative massive MIMO and cell-free massive MIMO systems. The both systems adopt the SIC-GAMP-base JADCE scheme \cite{Ke2020JADCE}.}}
\label{Fig14}
\end{figure}

For GFMA in cell-free massive MIMO systems, the APs transfer the pilot signals received from all the active devices to the related processing unit. Based on the MMV CS models in (\ref{Eq2}) and (\ref{Eq3}), the processing unit can perform JADCE by jointly processing the received signals from multiple APs using for example the SIC-GAMP algorithm \cite{Ke2020JADCE}. Table~\ref{Tab4} summarizes the JADCE schemes in cooperative massive MIMO systems. In Fig.~\ref{Fig14}, to verify the superiority of cell-free massive MIMO-based IoT networks, a conventional non-cooperative multi-cell massive MIMO architecture is compared  as the benchmark, where each BS only serves its own cell's devices and treats the inter-cell interference as noise. Here, we assume that $K = 2800$ devices are uniformly distributed in the network having a radius of $R_{\rm net} = 2.65$ km and $B = 7$ APs are geographically distributed to serve these devices. The AP-to-AP distance is $d = \sqrt{3}$ km, the number of active devices is $K_a = 140$, the number of AP antennas is $N_r = 16$, and the number of cooperating APs at each DPU is $N_{co}$. The SIC-GAMP-base JADCE scheme \cite{Ke2020JADCE} is employed by the both systems. As shown in Fig.~\ref{Fig14}, the cell-free massive MIMO network achieves much better AER and NMSE performance than the traditional non-cooperative multi-cell massive MIMO network architecture. It can also be seen that by increasing the number of APs for cooperation $N_{co}$, the performance of edge computing approaches that of cloud computing. In particular, we observe that only $N_{co} = 4$ APs are required for edge computing to obtain almost the same performance as cloud computing. This is because the signals received at the APs far away from a device are approximately zero due to the severe path loss, and incorporating them can hardly improve the JADCE performance further.

\begin{figure*}[!t]
\begin{center}
\subfloat[]{\includegraphics[width=13cm]{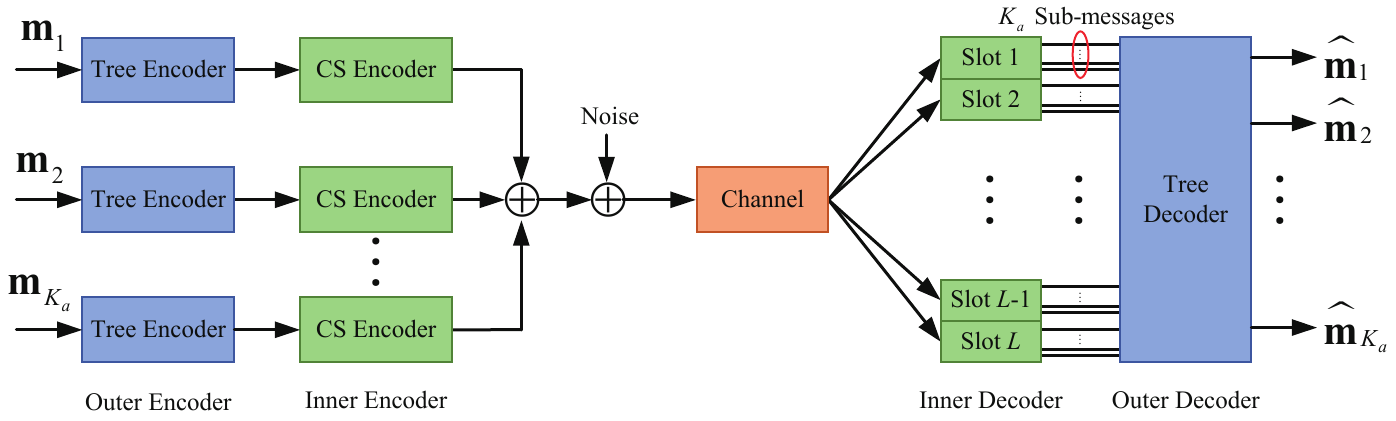}}
\hspace{5mm}
\subfloat[]{\includegraphics[width=13cm]{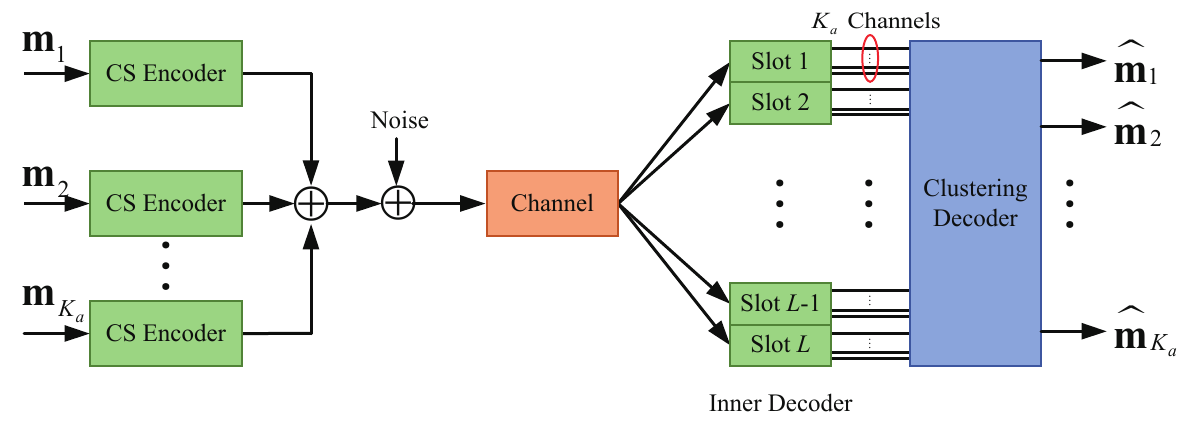}}
\end{center}
\caption{\small{CCND for GFMA: (a)~coupled CS-based transmission scheme, and (b)~uncoupled CS-based transmission scheme for massive MIMO systems \cite{Shyianov2021URA} \textcircled{c}IEEE.}}
\label{Fig15}
\end{figure*}

\subsection{Non-Coherent Data Detection}\label{S4.3}

The aforementioned GFMA schemes adopt a coherent data detection framework, where the detection performance is highly dependent on the accuracy of the CSI estimate. These solutions become inefficient or even impractical in high-mobility communications scenarios with small data packets, since the devices have to frequently transmit non-orthogonal pilot sequences for the CSI update. To address this limitation, two non-coherent detection frameworks were introduced, where the payload data of active devices is directly detected from the overlapped received signal, without any knowledge of the full CSI.

\subsubsection{Common Codebook-Based Non-Coherent Detection}\label{S4.3.1}

We first consider the unsourced random access scenarios, where the BS is solely interested in the list of transmitted messages without regard to their individual sources. In practice, the unsourced random access is motivated by the content-oriented IoT applications~\cite{Shao2020URA}. For example, in the quality inspection process of smart factories, many sensors are distributed at different positions on the production line to acquire the quality of products. The server only concerns about the weighted average of these sensors' quality information and does not have to know the identities of sensors that generate it. Focusing on unsourced random access, an efficient common codebook-based non-coherent detection (CCND) framework was developed in~\cite{Polyanskiy2017URA, Ordentlich2017URA, Vem2017URA, Marshakov2019URA, Tanc2021URA, Pradhan2020URA, Ahamdi2021URA, Truhachev2021URA, Pradhan2022URA}.

In the CCND framework, all the potential devices share a common codebook hardwired at the moment of production. When accessing the network, a specific active device $k$ first maps its $B$ information bits into an integer $b \in \left\{1, 2, \cdots, 2^B\right\}$, then the $b$th codeword ${\bf c}_b\in \mathbb{C}^{L\times 1}$ of the common codebook ${\bf C} = \left[{\bf c}_1, \cdots, {\bf c}_{2^B}\right] \in \mathbb{C}^{L \times 2^B}$ is directly transmitted to the BS, where $L$ is the length of codewords. Define a set of $K2^B$ Bernoulli random variables $\left\{\delta_{k,b} |  k = 1, \cdots, K; b = 1, \cdots, 2^B\right\}$ to model the device activity and codeword selection behavior. Specifically, $\delta_{k,b} = 1$ if the $k$th device is active and it selects the $b$th codeword to transmit; and $\delta_{k,b} = 0$ otherwise. Therefore, the overlapped received signal at the BS is expressed as
\vspace*{-1.5mm}
\begin{equation}\label{Eq4}
{\widehat {\bf Y}} = \sum_{k=1}^K \sum_{b=1}^{2^B} {\bf c}_b \delta_{k,b} {\bf h}_k^{\rm T} + {\widehat {\bf N}} = {\bf C}{\bm \Delta}{\bf H} + {\widehat {\bf N}},
\vspace*{-0.5mm}
\end{equation}  
where ${\bm \Delta}= \left[{\bm \delta}_1, {\bm \delta}_2, \cdots, {\bm \delta}_K\right] \in {\mathbb B}^{2^B \times K}$ is the codeword selection matrix with ${\bm \delta}_k = \left[\delta_{k,1}, \cdots, \delta_{k,2^B}\right]^{\rm T}$ $\in {\mathbb B}^{2^B \times 1}$, and ${\bf H} = \left[{\bf h}_1, {\bf h}_2, \cdots, {\bf h}_K\right]^{\rm T} \in \mathbb{C}^{K \times N_r}$ is the massive-access channel matrix. Note that the transmitted information is encoded in the non-zero indices of the codeword selection matrix ${\bm \Delta}$. Besides, ${\bm \Delta}$ contains only $K_a$ non-zero rows each of which having a single non-zero entry. Therefore, the data detection is equivalent to detecting the non-zero row indices of ${\bm \Delta}$ based on ${\widehat {\bf Y}}$ and the known ${\bf C}$, which can be formulated as a MMV CS problem of (\ref{Eq1}) by combining the codeword selection matrix and the massive-access channel matrix as a sparse matrix to be estimated, i.e., ${\bf X} = {\bm \Delta}{\bf H} \in \mathbb{C}^{2^B \times N_r}$ with $N = 2^B$ and $Q = N_r$, while defining the sensing matrix as ${\bf \Psi} = {\bf C}$ with $M = L$. In particular, each active device contributes a single non-zero coefficient in a specific column of ${\bf X}$, hereby resulting in a $K_a$-sparse $2^B$-dimensional vector. Moreover, different columns, i.e., different receive antennas, have a common sparsity pattern. This structured sparsity can be exploited for enhancing CS recovery performance \cite{Shao2020URA}.

\begin{table*}[t!] 
\caption{\small{Summary of CCND schemes for unsourced GFMA}}
\begin{center}
\resizebox{\linewidth}{!}{   
\begin{tabular}{|l|l|l|l|}
\hline
\multicolumn{1}{|c|}{Reference}										&		\multicolumn{1}{c|}{BS Architecture}		&		\multicolumn{1}{c|}{Advances} & { Complexity}\\ 
\hline
Y. Polyanshiy \cite{Polyanskiy2017URA}     			& 		Single-antenna                       		& 		The first work of a CCND framework for unsourced random access & {${\cal O}\left(LK2^B\right)$} \\ 
\hline
O. Ordentlich, \emph{et al.} \cite{Ordentlich2017URA}     			& 		Single-antenna                       		& 		The first low-complexity CCND scheme & {${\cal O}\left(2^{B}\right)$} \\ 
\hline
V. K. Amalladinne, \emph{et al.} \cite{Amalladinne2018URA} 			& 		Single-antenna                       		& 		A coupled CS-based CCND scheme for reduced complexity & {${\cal O}\left(L2^{B_{\rm sub}}\right)$} \\ 
\hline
A. Fengler, \emph{et al.} \cite{Fengler2021URA}       				& 		Single-antenna                       		& 		Leverage sparse regression code to reduce the codebook size & {${\cal O}\left(L2^{B_{\rm sub}}\right)$} \\ 
\hline
A. Fengler, \emph{et al.} \cite{Fengler2019URA}     				& 		Massive MIMO                         		& 		Extend the problem to the massive MIMO systems & {${\cal O}\left(LN_r2^{B_{\rm sub}}\right)$} \\ 
\hline
S. Haghighatshoar, \emph{et al.} \cite{Haghighatshoar2018URA} 		& 		Massive MIMO                         		& 		\begin{tabular}[c]{@{}l@{}} A covariance-based CS recovery algorithm to reduce the complexity of \\ inner decoder \end{tabular} & {${\cal O}\left(2^{B_{\rm sub}}\right)$} \\ 
\hline
A. Decurninge, \emph{et al.} \cite{Decurninge2021URA}    			& 		Massive MIMO                         		& 		A tensor-based transmission scheme for CCND in massive MIMO systems & {${\cal O}\left(2^{K_aB_{\rm sub}}\right)$} \\ 
\hline
V. Shyianov, \emph{et al.} \cite{Shyianov2021URA}      				& 		Massive MIMO                         		& 		An uncoupled CS-based CCND scheme for massive MIMO systems & {${\cal O}\left(K_a^3\right)$} \\ 
\hline
X. Xie, \emph{et al.} \cite{Xie2022URA}           					& 		Massive MIMO                         		& 		Leverage the diversity of devices' AoAs to resolve the collisions & {${\cal O}\left(LN_r2^{B_{\rm sub}}\right)$} \\ 
\hline
\end{tabular}}c
\end{center}
\label{Tab5}
Notes: $B_{\rm sub}$ is the length of sub-blocks.
\end{table*}

However, since the number of codewords scales exponentially with the message length, the computational complexity of the CS recovery is prohibitive even for the case with dozens of information bits. This becomes a major obstacle for practical implementation of CCND. To overcome this obstacle, the authors in \cite{Amalladinne2018URA} introduced a coupled CS transmission scheme for reduced complexity, as shown in Fig.~\ref{Fig15}\,(a). The scheme consists of outer and inner encoders/decoders. For the outer encoder, each active device's message is non-uniformly divided into multiple sub-messages and the redundancy check bits are added in these sub-messages to form the sub-blocks having uniform length. Then, the inner encoder is employed to map the sub-blocks to the codewords in a common codebook, which are transmitted in their corresponding slots. At the BS, the inner decoder first recovers the lists of sub-messages for all the slots and the sub-messages are then stitched together by the outer decoder using the redundancy check bits. Afterward, the detection performance is further enhanced by passing information between the CS-based inner decoder and the outer decoder dynamically \cite{Amalladinne2020URA}. Besides, the sparse regression code was introduced to reduce the size of codebook, where the sub-messages are encoded by the structured linear combination of the columns of the common codebook \cite{Fengler2021URA}.

The prior works \cite{Amalladinne2018URA, Fengler2021URA, Amalladinne2020URA} are limited to single-antenna systems. As a remedy, the authors in \cite{Fengler2019URA} investigated the CCND problem in massive MIMO systems and revealed that the transmit power per bit can be made arbitrary small if the number of BS antennas is sufficiently large. By equipping massive antennas at the BS, a low-complexity covariance-based CS recovery algorithm was developed for the implementation of an inner decoder \cite{Haghighatshoar2018URA}. Besides, a tensor-based transmission scheme was proposed for block fading channels in massive MIMO systems \cite{Decurninge2021URA}. In addition, the problem was extended to the cell-free massive MIMO systems, where the detection performance can be further improved by considering the cooperation of geographically distributed APs \cite{Shao2020URA}. 

Although the coupled CS-based schemes significantly reduce the decoding complexity, the spectral efficiency is scarified due to the employment of redundancy check bits. {To avoid the loss,} the authors in \cite{Shyianov2021URA} proposed that the strong correlation between the reconstructed MIMO channels in different slots provided enough information to stitch the sub-messages and devised an uncoupled CS-based scheme by removing the outer encoder and decoder. Specifically, the message is uniformly divided into multiple sub-messages, which are directly transmitted in slot-wise using the common codebook-based encoder, as illustrated in Fig.~\ref{Fig15}\,(b). At the BS, with the recovered sub-messages, an expectation maximization-based clustering algorithm is designed to obtain the original message. Since no redundancy is introduced, the spectral efficiency is dramatically improved. For uncouple CS-based schemes, however, due to the small number of information bits of sub-messages and the absence of check bits, the collision, i.e., multiple active devices select the same codeword, becomes a new challenge. To address this issue, the work \cite{Xie2022URA} exploited the diversity of different devices' angles of arrival (AoAs) to resolve the collision and leveraged the angular-domain sparsity of massive MIMO channels to improve the detection performance.  

A brief summary of the aforementioned CCND schemes are provided in Table~\ref{Tab5}.

\subsubsection{Individual Codebook-Based Non-Coherent Detection}\label{S4.3.2}

Despite of its many advantages, CCND is limited to unsourced random access scenarios. In practice, most IoT applications rely on sourced random access, where the server concerns both the transmitted messages and the identities (IDs) of active devices that generate them. By making several minor modifications to the packet structure, a straightforward solution is to embed the device ID in the transmitted information bits, and then to map the combined device ID and payload data onto a codeword in the common codebook \cite{Fengler2021URA2}. To identify $K$ devices, a device ID sequence with at least $\lceil {\rm log}_2\left(K\right) \rceil$ bits is embedded, where the operator $\lceil \cdot \rceil$ rounds a real number to the nearest integer larger or equal to it. Hence, the payload efficiency is significantly degraded, especially for the scenarios with massive number of devices and small data packets.

\begin{figure}[!t]
\begin{center}

+\includegraphics[width=1\columnwidth, keepaspectratio]{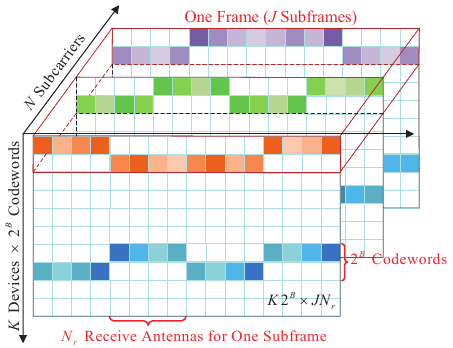}
\end{center}
\caption{\small{Space-time-frequency structured sparsity of individual codebook-based non-coherent detection for GFMA in massive MIMO-OFDM systems \cite{Qiao2022Blind} \textcircled{c}IEEE.}}
\label{Fig16}
\end{figure}

\begin{figure*}[!t]
\begin{center}
\subfloat[]{\includegraphics[width=0.85\columnwidth]{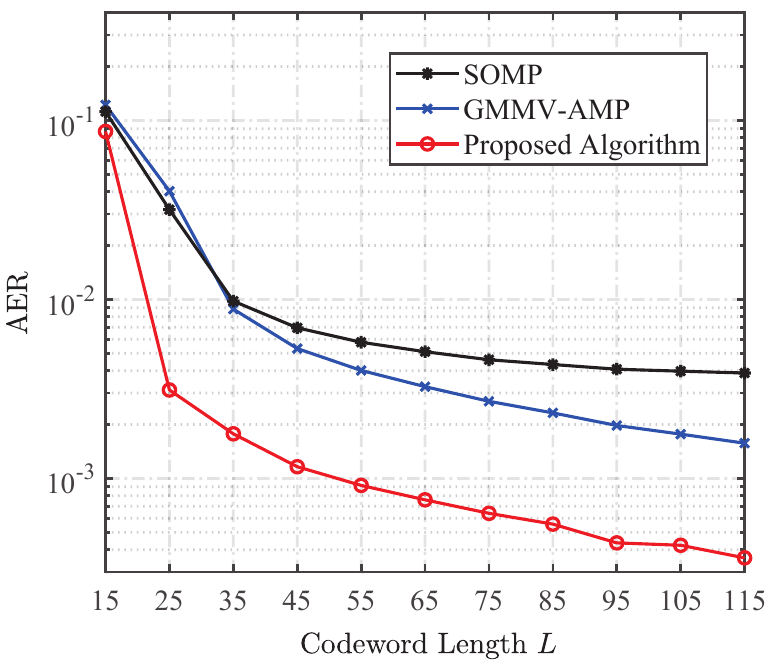}}
\hspace{5mm}
\subfloat[]{\includegraphics[width=0.85\columnwidth]{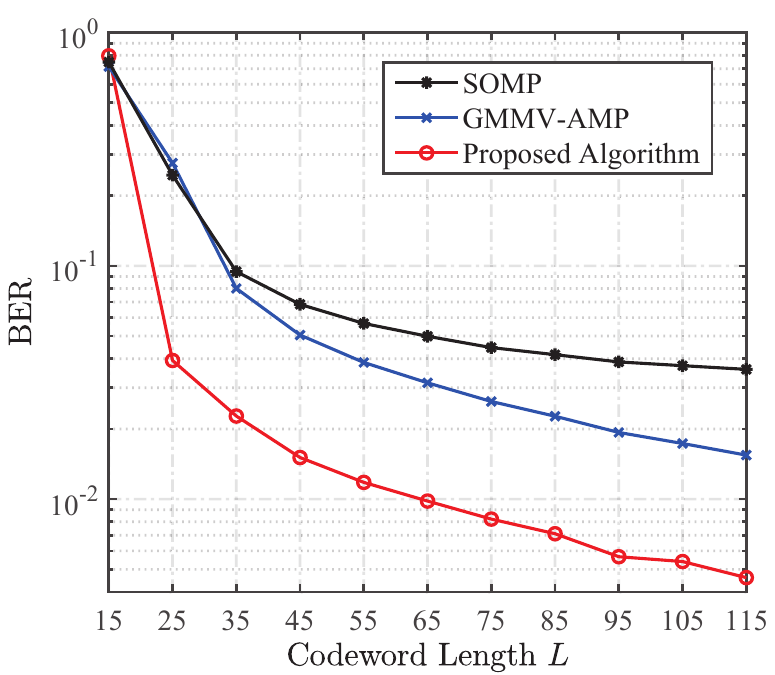}}
\end{center}
\caption{{(a)~AER, and (b)~BER performance comparison of individual codebook-based non-coherent detection schemes for GFMA in massive MIMO-OFDM systems \cite{Qiao2022Blind} \textcircled{c}IEEE.}}
\label{Fig17}
\vspace*{-6mm}
\end{figure*}

Recently, an individual codebook-based non-coherent detection scheme was proposed to support sourced GFMA more efficiently \cite{Qiao2022Blind}. In this scheme, each device is allocated with an individual codebook, i.e., ${\widetilde {\bf C}}_k = \left[{\tilde {\bf c}}_{k,1}, {\tilde {\bf c}}_{k,2}, \cdots, {\tilde {\bf c}}_{k,{2^B}}\right] \in \mathbb{C}^{L \times 2^B}$, to convey $B$-bit information, where $L$ is the length of codewords. Based on the $B$ information bits to be conveyed, each active device first select a codeword from its individual codebook, and then transmit the codeword on one subcarrier of $L$ successive OFDM symbols. In this context, a total of ${\widetilde N}2^B$ bits can be transmitted adopting OFDM with ${\widetilde N}$ subcarriers. Defining $L$ successive OFDM symbols as a subframe, the device activity during a frame of $J$ subframes and the CSI within a subframe are assumed to be invariant. At the BS, the signal received at the ${\tilde n}$-th subcarrier in the $j$-th subframe is expressed as
\vspace*{-2mm}
\begin{equation}\label{Eq5}
{\widetilde {\bf Y}}_{\tilde n}^j = \sum_{k=1}^K {\widetilde {\bf C}}_k{\widetilde {\bf X}}_{k,{\tilde n}}^j +  {\bf N}_{\tilde n}^j = {\widetilde {\bf C}}{\widetilde {\bf X}}_{\tilde n}^j + {\bf N}_{\tilde n}^j,
\end{equation}  
where ${\widetilde {\bf X}}_{k,{\tilde n}}^j = \alpha_k {\bf e}_{k,{\tilde n}}^j\left({\bf h}_{k,{\tilde n}}^j\right)^{\rm T} \in \mathbb{C}^{2^B \times N_r}$ is the equivalent channel matrix, $\alpha_k$, ${\bf e}_{k,{\tilde n}}^j \in {\mathbb B}^{2^B \times 1}$, and ${\bf h}_{k,{\tilde n}}^j \in {\mathbb C}^{N_r \times 1}$ are the activity indicator, codeword selection vector, and MIMO channel vector, respectively. Furthermore, ${\widetilde {\bf C}}=\left[{\widetilde {\bf C}}_1, \cdots, {\widetilde {\bf C}}_K\right] \in \mathbb{C}^{L \times K2^B}$ and ${\widetilde {\bf X}}_{\tilde n}^j=\left[\left({\widetilde {\bf X}}_{1,{\tilde n}}^j\right)^{\rm T}, \cdots ,\left({\widetilde {\bf X}}_{K,{\tilde n}}^j\right)^{\rm T}\right]^{\rm T} \in \mathbb{C}^{K2^B \times N_r}$. Combining ${\widetilde N}$ subcarriers and $J$ subframes, the non-coherent data detection can be formulated as a MMV CS problem of (\ref{Eq1}), where the measurement matrix is given as ${\bf Y} = \left[{\widetilde {\bf Y}}_1^1, \cdots, {\widetilde {\bf Y}}_{\widetilde N}^J\right] \in \mathbb{C}^{L \times J{\widetilde N}N_r}$ with $M = L$ and $Q = J{\widetilde N}N_r$, the sensing matrix is expressed as ${\bf \Psi} = {\widetilde {\bf C}}$ with $N = K2^B$, and the sparse channel matrix to be estimated is expressed as ${\bf X} = \left[{\widetilde {\bf X}}_{1}^j, \cdots, {\widetilde {\bf X}}_{\widetilde N}^J\right] \in \mathbb{C}^{K2^B \times J{\widetilde N}N_r}$. Here, both the device activity and transmitted information are encoded in the non-zero row indexes of ${\bf X}$. Moreover, ${\bf X}$ exhibits structured sparsity in the space, time, and frequency domains, as illustrated in Fig. \ref{Fig16}, where the common sparsity can be observed at different receive antennas and subcarriers within a subframe, and different subframes have an approximate common sparsity pattern. 

The authors in \cite{Qiao2022Blind} proposed an AMP-based joint activity and blind information detection algorithm to leverage the space-time-frequency structured sparsity of ${\bf X}$ for improving AER and BER performance. Fig. \ref{Fig17} demonstrates the superiority of the proposed algorithm \cite{Qiao2022Blind}, where the state-of-the-art SOMP algorithm \cite{Determe2017SOMP} and generalized MMV-AMP (GMMV-AMP) algorithm \cite{Ke2020JADCE} without taking the space-time-frequency structured sparsity into account are used as the benchmarks for comparison. Here, it is assumed that the number of devices is $K = 100$ with $K_a = 10$ active devices, ${\widetilde N} = 512$, $B = 1$, $J = 2$, and $N_r = 2$. It is clear that the proposed algorithm \cite{Qiao2022Blind} significantly outperforms the two benchmarks by fully exploiting the structured sparsity of the equivalent channel matrix.

\section{Future Research Directions}\label{S5}

Although extensive research efforts have been made to accelerate the development of the CS-based GFMA paradigm, numerous practical challenging issues still remain open to be resolved. In this section, we discuss some future research directions to address the key challenges in implementing the CS-based GFMA paradigm {for the 6G massive communication}.

%

\subsection{Practical Hardware Constraints}\label{S5.1}

{As reviewed in the previous sections, most GFMA schemes consider ideal hardware assumptions, such as fully digital MIMO architecture at the BS, infinite-resolution analog-to-digital converters (ADCs), perfect synchronization between the devices and the BS, equal amplitudes and phases between in-phase (I) and quadrature (Q) branches, i.e., I/Q balance, etc. Due to the resulting high hardware cost and huge power consumption, the GFMA in massive MIMO systems should be investigated under the more practical hybrid MIMO architecture \cite{Mendez-Rial2016Hybrid, Alkhateeb2014Hybrid} and low-resolution ADCs \cite{Fan2015ADC}. Moreover, due to the imperfect hardware, the carrier frequency offset caused by the asynchronization of the oscillators between the devices and the BS \cite{Sun2022CFO, Li2019CFO} as well as the I/Q imbalance \cite{Valkama2001IQ} should be further incorporated into the CS-based GFMA paradigm. Considering these practical hardware constraints, the problem formulations and receive algorithms presented in the previous sections are not directly applicable, and thus the corresponding GFMA schemes have to be redesigned.}

\subsection{GFMA in Space-Air-Ground-Sea Integrated Networks}\label{S5.2}

Most existing works usually implement various IoT applications in the terrestrial cellular networks. 
However, {the 6G ubiquitous connectivity is expected to rely on the space-air-ground-sea integrated networks (SAGEINs)}. Due to the inherent limitations of terrestrial infrastructures, it is impractical or uneconomic for deploying terrestrial BSs to seamlessly integrate the devices distributed across the ground, ocean, and air \cite{Qiu2019SAG}. As supplements to terrestrial networks, non-terrestrial networks (NTNs), including satellite constellations at different Earth orbits, high-altitude platform (HAP) networks, and unmanned aerial vehicle (UAV) networks, can offer effective coverage to remote areas where terrestrial BSs are unavailable. {To connect the different layers in hierarchical SAGEINs, an efficient GFMA scheme design is required, though} it involves numerous challenging issues, such as the seamless integration of heterogeneous networks \cite{Qiu2019SAG}, efficient cooperation between different networks \cite{Cheng2019SAG}, channel modeling for aerial communication links \cite{Khawaja2019chModel}, high ground-to-space path loss \cite{AL-Hourani2016PL}, etc. Therefore, considerable research efforts should be directed to address these challenging issues in the deployment of GFMA in SAGEINs.

\subsection{Deep Learning-Enhanced Design}\label{S5.3}

{Due to the remarkable accomplishments demonstrated by deep learning in various fields, such as computer vision, natural language processing, and image recognition, it is expected to serve as one of the primal driving forces to propel the advancement of 6G.} Recently, deep learning has showed its huge potentials in resource allocation, signal processing, channel estimation, and transceiver design for wireless communication systems \cite{Qing2020ML, Shao2021ML,Ma2023ML}. In particular, deep learning can fully leverage the implicit information in the available data and the benefits of well-developed wireless communication models, to reap a better performance or a lower complexity compared with the traditional design approaches. For GFMA in massive MIMO systems, the massive numbers of devices and BS antennas make the problem dimension extremely large, where the implementation of the aforementioned CS-based GFMA schemes would result in a high complexity. The deep learning-enhanced design is expected to provide a low-complexity and better-performance alternative. For example, the authors in \cite{Zhang2019DL} proposed a deep neural network (DNN)-aided message passing-based block sparse Bayesian learning algorithm to solve the JADCE problem of GFMA, which could approach the lower NMSE bound. Meanwhile, the authors in \cite{Kim2020DL} proposed a DNN-based activity detection scheme for GFMA. Considering the imperfect CSI, a DNN-based on variational autoencoder is further developed for  activity detection in GFMA \cite{Zhao2020DL}. In \cite{Yu2023DL}, the authors proposed a JADCE neural network, which fully exploits the information contained in the received preamble and data signals for improved performance. Moreover, the authors in \cite{Bai2022DL} developed a {machine} learning framework, where the information distilled from the initial data recovery phase are utilized to further enhance channel estimation, which in turn improves data recovery performance.

However, the analytical framework and the generalization of the deep learning-based approaches pose the new challenges in their practical implementation to GFMA in massive MIMO systems. To address these issues, the model-driven deep learning framework can be adopted, where the well-developed wireless communication models and signal processing techniques are exploited to design the training network, leaving only a few key parameters that need to be trained \cite{Qing2020ML, Zhu2021DL}.

\subsection{GFMA for High-Speed IoT Applications}\label{S5.4}

Most existing works assume that the device activity and CSI remain unchanged during the considered time interval, e.g., \cite{Liu2018JADCE3, Shao2018JADCE, Chen2018JADCE, Shao2020JADCE, Jiang2021JADCE}. In addition, the devices are assumed to be perfectly synchronized. To support mobile IoT applications, such as smart traffic {and UAV communications in 6G}, a more complicated massive connectivity scenario should be further investigated, where the devices are moving at a high speed and thus result in fast time-varying channels. Furthermore, the devices can randomly access or leave the network, which leads to the time-varying device activity and the asynchronous transmission between the devices \cite{Lin2020Asy, Cheng2021TC}. In this context, the existing CS-based GFMA schemes will fail to work, and the transceiver should be redesigned to capture the variation of the channels and to handle the asynchronous transmission problem. Here, the temporal correlation of the channels can be exploited for improving multi-device detection performance \cite{Cheng2021TC}.
  
\subsection{Joint Activity Detection, {Channel} Estimation, and Data Decoding}\label{S5.5}

Most previous works  assumed a two-phase processing, i.e., JADCE and data decoding. There is another line of research, i.e., joint activity detection, channel estimation, and data decoding (JADCEDD), where partially detected data can be used as soft pilots to enhance the channel estimation accuracy in
an iterative fashion. In the field, the authors of \cite{Li2022JADCEDD} proposed a belief propagation based-joint device detection, channel estimation and data decoding algorithm for unsourced massive access, in which the CE results can be enhanced by regarding the corrected decoded data in the LDPC phase as extra pilots to execute the second CE. Similarly, in \cite{Bian2023JADCEDD}, the authors achieve the JADCEDD by BiG-AMP algorithm in a turbo receiver that can effectively exploit the common sparsity pattern in the received pilot and data signal, and improve the data detection performance by incorporating with channel decoder. Furthermore, the authors of \cite{Renna2023JADCEDD} proposed a bilinear message-scheduling GAMP algorithm for JADCEDD in a grant-free massive MIMO scenario. By applying the activity detection results or the residual for the message to determine the update of messages, they introduced two message-scheduling techniques to reduce the computational cost while maintaining the detection performance. Besides, the authors in \cite{Zhou2023JADCEDD} extend this topic to the NOMA-OTFS system in LEO satellite IoT, achieving good detection performance by overcoming the long round-trip latency and severe Doppler effect. In, general, the JADCEDD is more practical than the JADD and usually exhibits better performance than the JADCE. However, this advantage is obtained at the expense of computational complexity, such as introducing more iterations. How to strike a trade off is needed to be considered in the future research.

\section{Conclusions}\label{S6}

The future 6G massive communication is expected to require instant and seamless wireless connectivity for extremely large numbers of devices, which is a key enabler of the digital transformation of many aspects of society. Thanks to the recently completed infrastructures and well-developed technologies, the existing cellular networks can serve as a solid foundation for implementing massive connectivity in practice. This review has explored various typical IoT use cases and their service requirements. Moreover, the state-of-the-art IoT standards and the random access solutions from the both industry and academic communities have been reviewed. In particular, we have pointed out the limitations of the existing random access solutions, which do not take into account the inherent sparse communication behavior of massive communication. Against this background, a CS-based GFMA paradigm has been introduced, where the active devices directly access the network without any scheduling, and the activity detection, channel estimation, and/or data detection at the BS can be formulated as an SMV/MMV CS problem. Under the CS-based GFMA paradigm, various network architectures, transmission schemes, data detection frameworks, and receive algorithms can be flexibly incorporated to meet different service requirements of heterogeneous IoT applications. In this respect, we have detailed the roadmap with evolutions from single-antenna to large-scale antenna array-based BSs, from single-station to cooperative massive MIMO systems, and from unsourced to sourced random access scenarios. Finally, we have discussed the key challenges and open issues to provide enlightening guidance for future research directions.

\vspace{30mm}

{\section*{Appendix}

A list of major abbreviations used in this paper is provided in Table VI.

\begin{table}[h]
\caption*{TABLE VI: List of Major Abbreviations}
\begin{tabular}{l|l}

%

\hline
\textbf{Abbreviations} & \textbf{Meanings} \\


%
%

\hline
ADC & Analog-to-digital converter \\

\hline
AER & Activity error rate \\

\hline
AMP & Approximate message passing \\

\hline
AoA & Angle of arrival \\

\hline
AP & Access point \\



\hline
CCND & Common Codebook-based non-coherent detection \\

\hline
CDMA & Code-domain multiple access \\

\hline
CIR & Channel impulse response \\

\hline
CPU & Central processing unit \\

\hline
C-RAN & Cloud radio access network \\

\hline
CS    & Compressive sensing       \\ 

\hline
CSI & Channel state information \\

\hline
DMRS & Demodulation reference signal \\


\hline
DPU & Distributed processing unit \\

\hline
DS & Dense spreading \\

\hline
DS-AMP & Doubly structured AMP \\

\hline
EC-GSM    &    \begin{tabular}[l]{@{}l@{}}  Extended coverage global system for \\ mobile communications \end{tabular} \\

\hline
EDT & Early data transmission \\

\hline
eDRX & Extended discontinuous reception \\

\hline
EM & Expectation maximization \\

\hline
FSRA & Four-step random access \\

\hline
GAMP & Generalized AMP \\

\hline
GFMA   & Grant-free massive access       \\ 

\hline
GMMV-AMP & Generalized multiple measurement vector AMP \\

\hline
GPRS   & General packet radio service \\


\hline
JADCE & Joint activity detection and channel estimation \\

\hline
JADD & Joint activity and data detection \\

\hline
LASSO & Least absolute shrinkage and selection operator \\

\hline
LDS & Low-density spreading \\

\hline
LPWAN & Low-power wide-area network   \\


\hline
MBM & Media-based modulation \\

\hline
MIMO & Multi-input multi-output        \\ 

\hline
mMTC & Massive machine-type communications        \\ 

\hline
MPA & Message passing algorithm \\

\hline
MUD & Multiuser detection \\

\hline
MUSA & Multiuser shared access \\

\hline
MUT & Multiuser transmit \\

\hline
NB-IoT &  Narrow-band Internet-of-Things \\

\hline
NOMA    & Non-orthogonal multiple access       \\ 



\hline
OAMP-ASL & Orthogonal AMP with accurate structure learning \\

\hline
OMA & Orthogonal multiple access \\

\hline
OMP & Orthogonal matching pursuit \\

\hline 
PIA-ASP & Prior-information aided adaptive subspace pursuit \\

\hline
PIA-MSMP &  \begin{tabular}[l]{@{}l@{}} Prior-information aided adaptive media modulation  \\ subspace matching pursuit \end{tabular} \\

\hline
PRACH & Physical random access channel \\

\hline
PSM & Power-saving mode \\

\hline
PUR & Pre-configured uplink resources \\

\hline
PUSCH & Physical uplink shared channel \\


\hline
RFID   & Radio frequency identification \\

\hline
GAGEIN & Space-air-ground-sea integrated network \\

\hline
SCMA & Sparse code multiple access \\

\hline
SE & State evolution \\



\hline
SISD & Structured iterative support detection \\

\hline
SIC-SSP & Successive inference cancellation based structured SP \\

\hline
SM & Spatial modulation \\

\hline
SMV & Single measurement vector \\

\hline
SP & Subspace pursuit \\

\hline
TLSS & Two-level structured sparsity \\

\hline
TSRA & Two-step random access \\
\hline

\end{tabular}
\end{table}

\end{document}